\def\notodocomments{}

\PassOptionsToPackage{dvipsnames,svgnames,x11names}{xcolor}
\documentclass[acmsmall,screen,review,nonacm]{acmart}
\AtBeginDocument{%
  \providecommand\BibTeX{{%
    \normalfont B\kern-0.5em{\scshape i\kern-0.25em b}\kern-0.8em\TeX}}}

\usepackage{tabularx}
\usepackage{booktabs}
\usepackage{wrapfig}
\usepackage{tcolorbox}

\usepackage{relsize}

\usepackage{xspace}

\usepackage{enumitem}

\usepackage{xcolor}

\definecolor{keywordcolor}{HTML}{0000FF}
\definecolor{typecolor}{HTML}{007070}
\definecolor{functioncolor}{HTML}{7928a1}
\definecolor{stringcolor}{HTML}{A31515}
\definecolor{commentcolor}{HTML}{6A9955}
\definecolor{numbercolor}{HTML}{098658}
\definecolor{annotationcolor}{HTML}{9A6700}

\usepackage{listings}
\usepackage[scaled]{beramono}

 \usepackage{tikz}
 \usepackage[edges]{forest}
 \usetikzlibrary{shapes.geometric, arrows.meta, positioning, fit}
\usepackage{setspace,caption}
\usepackage{flushend} %

\usepackage{url} %

\usepackage{cleveref}

\begin{document}

\newcommand{\Oca}{Oca\xspace} %

\newcommand{\conferencePageLimit}{18}

\newcommand{\todo}[1]{{\color{red}\bfseries [[#1]]}}
\ifdefined\notodocomments
  \renewcommand{\todo}[1]{\relax}
\fi

\newcommand{\martin}[1]{\todo{MK: #1}}

\newif\iflongversion
\ifdefined\createlongversion
  \longversiontrue
\fi
\newcommand{\longversion}[1]{\iflongversion #1\else\relax\fi}

\newif\ifanonymous
\anonymoustrue

\newcommand{\anonurl}[1]{\ifanonymous URL removed for anonymity.\else\url{#1}\fi}
\newcommand{\footnoteanonurl}[1]{\footnote{\anonurl{#1}}}

\def\|#1|{\mathid{#1}}
\newcommand{\mathid}[1]{\ensuremath{\mathit{#1}}}
\def\<#1>{\codeid{#1}}
\protected\def\codeid#1{\ifmmode{\mbox{\smaller\ttfamily{#1}}}\else{\smaller\ttfamily #1}\fi}
\newlist{researchquestions}{enumerate}{1}
\setlist[researchquestions]{label*=\textbf{RQ\arabic*}}

\newcommand{\CalledMethodsBottom}{\<@Call\-ed\-Meth\-ods\-Bottom>\xspace}
\newcommand{\CalledMethods}{\<@Call\-ed\-Meth\-ods>\xspace}
\newcommand{\EnsuresCalledMethods}{\<@En\-sures\-Call\-ed\-Meth\-ods>\xspace}
\newcommand{\MustCall}{\codeid{@Must\-Call}\xspace}
\newcommand{\MustCallAlias}{\codeid{@Must\-Call\-Alias}\xspace}
\newcommand{\MustCallUnknown}{\codeid{@Must\-Call\-Unknown}\xspace}
\newcommand{\CreatesMustCallFor}{\<@Creates\-Must\-Call\-For>\xspace}
\newcommand{\ResetMustCall}{\CreatesMustCallFor}

\newcommand{\trule}[2]{\[\frac{#1}{#2}\]}
\newcommand{\truleinline}[2]{\ensuremath{#1\mathrel{\vdash}#2}}
\newcommand{\hastype}[1]{\mathbin{:}\trtext{#1}}
\newcommand{\trcode}[1]{\codeid{\smaller\smaller #1}}
\newcommand{\trtext}[1]{\mbox{\smaller\smaller #1}}
\newcommand{\trquoted}[1]{\trcode{"}#1\trcode{"}}

\newcommand{\numTypeSystems}{11\xspace} %
\newcommand{\numModifiedTypeSystems}{2\xspace} %
\newcommand{\numProjects}{12\xspace}
\newcommand{\numLOC}{88,680\xspace}
\newcommand{\numHumanAnnos}{803\xspace}
\newcommand{\percentInferred}{39\todo{check}\%\xspace}
\newcommand{\warningReductionPercent}{45\todo{check}\%\xspace}
\newcommand{\tsSpecificLoC}{61\todo{check}\xspace}

\hyphenation{type-state}        %
\hyphenation{null-able}         %

\setlength{\leftmargini}{.75\leftmargini}
\setlength{\leftmarginii}{.75\leftmarginii}
\setlength{\leftmarginiii}{.75\leftmarginiii}

\newcommand{\prefigcaption}{\vspace{-5pt}}
\newcommand{\posttablecaption}{\vspace{-5pt}}

\addtolength{\textfloatsep}{-.25\textfloatsep}
\addtolength{\dbltextfloatsep}{-.25\dbltextfloatsep}
\addtolength{\floatsep}{-.25\floatsep}
\addtolength{\dblfloatsep}{-.25\dblfloatsep}

\newcommand{\zph}{\phantom{0}}
\newcommand{\zzph}{\phantom{00}}

\newcommand{\ie}{i.e.,\xspace}
\newcommand{\eg}{e.g.,\xspace}

\newcommand{\CollectionToObjectConverter}{\codeid{Col\-lec\-tion\-To\-Ob\-ject\-Con\-ver\-ter}\xspace}
\newcommand{\sourceTypeIsAssignableToTargetType}{\codeid{source\-Type\-.is\-Assign\-able\-To(\-tar\-get\-Type)}}

\title{Template-Free Dynamic Invariant Detection}

\author{Erfan Arvan}
\email{ea442@njit.edu}
\orcid{0000-0002-6170-1957}
\affiliation{%
  \institution{NJIT}
  \city{Newark, NJ}
  \country{USA}
}

\author{Michael D. Ernst}
\affiliation{%
  \institution{UW CSE}
  \city{Seattle, WA}
  \country{USA}}
\email{mernst@cs.washington.edu}
\orcid{0000-0001-9379-277X}
\orcid{0000-0001-9379-277X}

\author{Martin Kellogg}
\affiliation{%
  \institution{NJIT}
  \city{Newark, NJ}
  \country{USA}
}
\email{martin.kellogg@njit.edu}
\orcid{0000-0002-3185-2340}
\begin{abstract}
A dynamic invariant detector proposes
invariants, such as method pre- and postconditions, and
rules out those that
do not hold at run time.
Traditional detectors propose invariants from a
fixed library of templates, which limits their expressiveness.
A large language model (LLM) can propose invariants instead of, or in
addition to, template-derived ones.
With the right prompt and context choices, we found that
this idea can both improve the invariants' expressiveness
and reduce the cost of
inferring the invariants.

Greater expressiveness brings complications: an LLM-proposed
invariant can execute arbitrary code \emph{during invariant
evaluation}, causing
side effects or infinite loops.  Our novel
instrumentation strategies mitigate these problems.
We wrap each candidate and evaluate only one invariant
per thread at a time, so nested invariant evaluations
cannot trigger one another indefinitely.  To recover from non-nested hangs, we
monitor each run for stalls and retry it under a growing time budget.
Because we cannot prove a candidate side-effect-free in advance, we
use delta debugging over the invariants evaluated during a failing
test to isolate and disable a side-effecting candidate
without discarding the rest of the run.

We compared our implementation, \Oca, to the popular template-based Daikon
invariant detector on real open-source Java projects. \Oca
produces more expressive invariants: on average, 28\% of its held
invariants call a project-specific method, which Daikon's templates
cannot express. Its invariants are more often valid, with a
false-positive rate of 18\% in a manual inspection, \emph{vs.}\ 76\%
for Daikon, and more useful: as properties for property-based testing,
they exposed 9 of 25 real bugs, compared with 4 for Daikon. \Oca also
costs less, using 566 minutes and 27\,GB of disk,
compared with 2{,}817 minutes and 827\,GB for Daikon.

\label{dummy-label-for-etags:abstract}
\end{abstract}

\maketitle
\thispagestyle{plain}
\pagestyle{plain}

\section{Introduction}
\label{sec:intro}

\begin{figure}[t]
\centering
\begin{tikzpicture}[
  node distance=6pt,
  box/.style={rectangle, rounded corners=4pt, draw, text width=4.4cm, minimum height=0.95cm, align=center, font=\scriptsize},
  genbox/.style={box, fill=orange!12, draw=orange!60},
  checkbox/.style={box, fill=cyan!10, draw=cyan!55!black!70},
  graybox/.style={box, fill=gray!10, draw=gray!50},
  clabel/.style={font=\tiny, text=gray},
  head/.style={font=\small\bfseries},
  arr/.style={-{Stealth[length=4pt]}, gray},
  darr/.style={-{Stealth[length=4pt]}, gray, dashed},
]
\node[graybox] (scan)    {Prepare \& scan\\[-2pt]{\tiny\color{gray}Copy source; identify program points}};
\node[genbox, below=of scan]   (llm)     {LLM query\\[-2pt]{\tiny\color{gray}Context + prompt per program point}};
\node[checkbox, below=of llm]      (filter)  {Filter, deduplicate \& compile\\[-2pt]{\tiny\color{gray}Reject invalid expressions; inject candidates as probes, recompiling iteratively}};
\node[checkbox, below=of filter]   (exec)    {Execute test suite\\[-2pt]{\tiny\color{gray}Suppress nesting; recover from hangs; isolate test-exposed side effects}};
\node[graybox, below=of exec]     (report)  {Report held invariants\\[-2pt]{\tiny\color{gray}Executed and never falsified}};

\draw[arr] (scan)   -- (llm);
\draw[arr] (llm)    -- (filter);
\draw[arr] (filter) -- (exec);
\draw[arr] (exec)   -- (report);

\draw[darr] (filter.south west) to[out=180,in=180,looseness=3] node[midway, clabel, left, align=right]{remove \&\\recompile\\until no error} (filter.north west);
\draw[darr] (exec.south west)   to[out=180,in=180,looseness=3] node[midway, clabel, left, align=right]{disable \&\\rerun until\\termination} (exec.north west);

\node[clabel, right=4pt of llm]    {C1};
\node[clabel, right=4pt of filter] {C1};
\node[clabel, right=4pt of exec, align=left]   {C2.1--\\C2.3};

\node[graybox, right=1.0cm of scan]     (dscan)   {Prepare \& compile\\[-2pt]{\tiny\color{gray}Copy source; compile program}};
\node[checkbox, right=1.0cm of llm]    (dinst)   {Instrument \& execute with Chicory\\[-2pt]{\tiny\color{gray}Instrument bytecode; run test suite and record runtime values}};
\node[genbox, right=1.0cm of filter] (dexec)   {Instantiate predefined invariants\\[-2pt]{\tiny\color{gray}\texttt{x = c}, \texttt{x < y}, \texttt{x != null}, \ldots}};
\node[checkbox, right=1.0cm of exec]   (dinfer)  {Filter \& suppress\\[-2pt]{\tiny\color{gray}Remove redundant and statistically unjustified invariants}};
\node[graybox, right=1.0cm of report]   (dreport) {Report held invariants\\[-2pt]{\tiny\color{gray}Executed and never falsified}};

\draw[arr] (dscan)  -- (dinst);
\draw[arr] (dinst)  -- (dexec);
\draw[arr] (dexec)  -- (dinfer);
\draw[arr] (dinfer) -- (dreport);

\draw[gray] ([xshift=4pt]dexec.north east) -- ([xshift=7pt]dexec.north east)
  -- node[clabel, midway, rotate=90, anchor=north]{offline} ([xshift=7pt]dinfer.south east)
  -- ([xshift=4pt]dinfer.south east);

\node[head, above=4pt of scan]  {\Oca};
\node[head, above=4pt of dscan] {Daikon};
\end{tikzpicture}
\caption{The pipelines of \Oca (left) and Daikon (right).}
\label{fig:pipeline}
\end{figure}
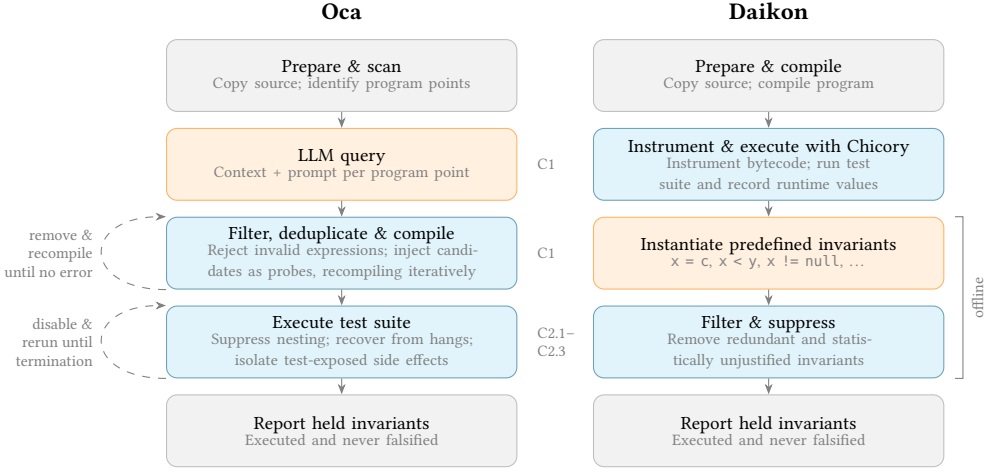

Invariants capture essential semantic properties of programs, such as
preconditions, postconditions, and design intent.
They support program understanding and serve as a foundation
for testing, verification, and debugging~\cite{ErnstPGMPTX2007}.
A dynamic invariant detection tool infers invariants
from executions.
Traditionally, the tool proposes
invariants from a fixed template library.  This can express only properties
that match one of its predefined syntactic forms: it cannot
represent invariants that depend on project-specific logic or domain
abstractions, and it
restricts relations to a small, fixed number of variables. These
limits are not implementation details of any one tool; they are structural
consequences of deriving candidates from a fixed grammar, whatever
that grammar is.
Daikon~\cite{ErnstPGMPTX2007} (\cref{fig:pipeline}, right) is the
most widely used tool of this design, in both research and
practice~\cite{AlonsoSR2023,MaLLXB2025,MolinaDPRAF2019,XiaSPDRR2025}.

Recent advances in large language models (LLMs) provide an
opportunity to move beyond fixed invariant templates. When supplied
with source code as context, an LLM can propose candidate invariants
that reflect program-specific constraints.

This paper introduces \Oca\footnote{\Oca is a South American root
vegetable, chosen as a nod to Daikon, which was named after another
root vegetable.}, a dynamic invariant detector realizing this
idea. Like Daikon, \Oca validates candidates against program
executions; unlike Daikon, it obtains them from an LLM rather than a
template grammar. For each method entry and exit, \Oca asks an LLM for
candidate invariants, filters them, injects the survivors into the
program source as instrumentation probes, and runs the existing test
suite (\cref{fig:pipeline}, left). A candidate that executes at least once
and is never falsified is reported as held.
Daikon by default checks invariants offline, against an execution trace
written by its Java frontend, Chicory. That is infeasible for arbitrary
expressions, which may depend on any part of the program state; %
checking candidates in place is also simpler and faster. Further
speedups come from evaluating only a comparatively small set of
LLM-proposed candidates.

\Oca also dispenses with the complex machinery Daikon uses to avoid
computing and reporting uninformative invariants. For example, an
invariant over unrelated quantities, such as \<myBoolean <= myPointer>,
should not be reported, so Daikon computes abstract types
\cite{Baker90,GuoPME2006} to determine which variables may meaningfully
be compared. Daikon also groups variables whose values are always equal,
so that each template is instantiated once per group rather than once
per variable; streamlines the common case of quickly falsified
invariants; and applies statistical tests to suppress invariants that
hold only by coincidence.
An advantage of checking a small, targeted set of potential invariants is
that \Oca needs none of these complexities.

Using an LLM to propose invariants raises two challenges.
The first (C1) is obtaining useful, well-formed candidates from the
LLM: without careful prompt design and context, models produce
ill-formed, vacuous, or irrelevant candidates, and even under a good
prompt, some candidates are syntactically invalid or fail to compile. The
second (C2) is ensuring that
invariants evaluated at run time do not corrupt the
inference process itself: they may
have side effects that perturb program state (C2.1), pass
otherwise-unseen values to the program's own routines{\,---\,}causing a
called method's own injected invariants to be evaluated against inputs
that arise only from invariant evaluation (C2.2){\,---\,}or fail to terminate
(C2.3).

We address these challenges as follows. For C1, we
empirically evaluated multiple prompt strategies and context
configurations, deriving the configurations from a
literature review of the context that LLM-based software engineering
techniques supply to the model (\cref{sec:config-selection}). \Oca
also filters ill-formed expressions before instrumentation, using
both syntactic checks and compilation (\cref{sec:filtering}). For side effects (C2.1), \Oca relies on the
project's test suite, using delta debugging to isolate and disable the
specific candidate invariant responsible for an observed test failure
(\cref{sec:side-effects}). For C2.2, \Oca suppresses nested
invariant evaluation when another check is already active on the
same thread (\cref{sec:nesting}). For nontermination (C2.3), the same re-entrancy check
rules out the nontermination that mutually triggering probes would
cause, and a background monitor detects
and disables suspected nonterminating candidates at the per-invariant
level, without discarding results for other invariants
(\cref{sec:hangs}).

We evaluated \Oca against Daikon on real open-source Java projects
along four complementary dimensions, and found that \Oca produces more
expressive, more valid, and more useful invariants at lower cost.
First, we compared the expressiveness of the inferred invariants
(\cref{sec:expressiveness-comparison}): on average, 28\% of \Oca's held
invariants call a project-specific method and 5\% relate more than
three variables; Daikon's templates can express neither.
Second, we manually checked both tools' output for false positives,
invariants that hold for the given executions but not in general
(\cref{sec:false-positive-analysis}): 18\% of sampled \Oca invariants
were false positives, compared with 76\% for Daikon.
Third, we compared the invariants as properties for property-based
testing (\cref{sec:usefulness}): \Oca exposed 9 of 25 real bugs,
compared with 4 for Daikon.
Fourth, we compared the tools' cost (\cref{sec:performance}): \Oca used
566 minutes and 27\,GB of disk in total, compared with 2{,}817 minutes
and 827\,GB for Daikon.
We deliberately adopt conservative evaluation choices throughout to
avoid overstating the benefits of LLM-generated invariants.
In summary, this paper makes the following contributions:
\begin{itemize}
    \item A characterization of the research challenges in
    using large language models for dynamic invariant
    detection.
    \item A language-agnostic design for template-free dynamic
    invariant detection: LLM-proposed candidates are checked in place
    by probes, with test-driven isolation of side effects, suppression
    of nested evaluation, and recovery from hangs.
    \item A literature review of the program context that LLM-based
    software engineering techniques supply to the model: a taxonomy
    derived by coding 181 papers from the latest editions of FSE,
    ICSE, ASE, and ISSTA, with the prevalence of each context type.
    \item An empirical study of prompt and context
    strategies for LLM-based invariant generation.
    \item \Oca, an implementation of our template-free invariant detector design for Java that
    supports both cloud-hosted and locally hosted LLMs.
    \item An empirical evaluation
    comparing \Oca and Daikon for expressiveness,
    validity, downstream usefulness, and inference cost.
\end{itemize}

\section{Technical Overview}
\label{sec:overview}

This section introduces the terminology used throughout the paper,
walks through \Oca's pipeline (\cref{fig:pipeline}) using a running example,
and, at each stage, states the research challenge that stage addresses
and the intuition behind \Oca's solution.
\Cref{sec:config-selection,sec:checking} treat each challenge in depth.

\subsection{Terminology}

In this paper,
a \textbf{program point} is a method
entry or exit; together they capture a method's
preconditions and postconditions.
A \textbf{candidate invariant} is an arbitrary Boolean
expression proposed for a program point; it may reference formal
parameters, \codeid{result} (at non-void method exits), and fields, and it may invoke
methods.
\Oca injects each candidate and executes the project's test suite;
each evaluation of a candidate at its program point is a \textbf{check}.
A candidate is reported as \textbf{held} if it compiles, is evaluated at
least once, and is never observed to be false; it is \textbf{falsified} if
any evaluation returns false or throws an exception; otherwise, it is
\textbf{never executed}.
We call the wrapper injected at the corresponding program point to evaluate the
candidate and record the outcome an \textbf{instrumentation probe}, or
simply a \textbf{probe} (\cref{sec:probe-structure} details its structure).
\Oca \textbf{disables} a candidate, leaving its probes in place but no longer
checking it, once it is falsified, causes a test failure, or hangs.

\begin{figure}[t]
\centering
\begin{lstlisting}[basicstyle=\tiny\ttfamily]
@Override
public @Nullable Object convert(@Nullable Object source, TypeDescriptor sourceType, TypeDescriptor targetType) {
    if (source == null) {
        return null;
    }
    if (sourceType.isAssignableTo(targetType)) {
        return source;
    }
    Collection<?> sourceCollection = (Collection<?>) source;
    if (sourceCollection.isEmpty()) {
        return null;
    }
    Object firstElement = sourceCollection.iterator().next();
    return this.conversionService.convert(firstElement, sourceType.elementTypeDescriptor(firstElement), targetType);
}
\end{lstlisting}
\caption{\CollectionToObjectConverter\codeid{.con\-vert} from Spring's~\cite{spring}
type-conversion infrastructure.}
\label{fig:running-example}
\end{figure}

\subsection{Running Example}
\label{sec:running-example}

The \codeid{convert} method of \cref{fig:running-example} takes
a source collection \codeid{source} of type \codeid{sourceType} and
produces a value of \codeid{targetType}.
If \codeid{sourceType} is already
assignable to \codeid{targetType}, the method returns \codeid{source}
unchanged.
\Oca reports this property at \<convert>'s exit:
\newline
{\lstset{keywordstyle=[3]{\normalcolor}}
\begin{lstlisting}[frame=tb,aboveskip=3mm,belowskip=3mm]
!sourceType.isAssignableTo(targetType) || result == source
\end{lstlisting}}

\noindent
By contrast, Daikon reports only
that \codeid{this.conversionService} is not side-effected.
Daikon's output says nothing about the method's semantics.
Inferring the key invariant, and doing so safely, requires \Oca to
address two challenges:
proposing invariants (C1) and evaluating candidates safely (C2).

\subsection{Proposing Invariants (C1)}
\label{sec:overview-proposal}

In initial experiments, a naively prompted LLM often proposed low-quality
invariants: tautologies such as
\sourceTypeIsAssignableToTargetType\codeid{ || !}\sourceTypeIsAssignableToTargetType;
trivial invariants such as
\codeid{source == null || source instanceof Object}, which merely reflects the
declared type of \codeid{source}; and false candidates such as
\codeid{result == source}, which does not always hold.

LLMs are also sensitive to the source-level context supplied in the prompt:
too little context can lead to generic, irrelevant, code-inconsistent, or
false candidates, while excessive or extraneous context can distract the model,
increase cost, and reduce the quality of its proposals. Prompt strategy and context
configuration together determine the quality of the candidates the LLM
proposes.

We address both facets of the proposal problem empirically rather than by
fixing design choices in advance: in \cref{sec:config-selection}, we evaluate a set of
prompt strategies drawn from prior work and select the best-performing
one, then evaluate and select a context configuration under that strategy.
All other evaluations use the selected combination,
\textbf{Few-Shot} with \textbf{Class-Aware} context.

\subsubsection{Filtering and Compiling Candidates}
\label{sec:overview-filtering}

Even with a well-designed prompt, a candidate is not guaranteed
to be high-quality, syntactically valid, or compilable:
it may be a tautology,
reference an identifier that is unavailable at the program point (\eg{} \codeid{firstElement != null}
at method entry, since \codeid{firstElement} is a local variable
defined only partway through the method body), violate a typing constraint
(\eg{} \codeid{source.size() > 0}, since \codeid{source} is statically typed
as \codeid{Object}), or be malformed
(\eg{} \sourceTypeIsAssignableToTargetType\codeid{ \&\& (result == source},
which is missing a closing parenthesis) (C1).
\Oca first applies cheap static filters (\cref{sec:static_filtering}),
which catch many problematic ones.
For the rest, the Java compiler provides a stronger check (\cref{sec:compilation}).

\subsection{Evaluating Candidates Safely (C2)}
\label{sec:overview-safety}

\subsubsection{Side Effects (C2.1)}

Evaluating a candidate can cause side effects that corrupt program state.
A side effect can make a valid invariant appear falsified or a falsified
invariant appear to hold.

\Oca does not attempt to prove a candidate side-effect-free before running it.
Precisely determining side-effect freedom is undecidable in general, and the
purity-analysis tools available in practice rely on conservative
approximations that classify many side-effect-free methods as potentially
side-effecting~\cite{helm2018unified}. Filtering candidates on the results of
such an analysis would wrongly reject many useful candidates that invoke
side-effect-free methods.

\Oca derives a more practical but weaker guarantee from the project's test suite.
The intuition is that if a side effect occurs,
it is likely to change the tests' behavior.
If a test fails, \Oca uses delta debugging to find the invariants that
caused that failure and disables them in the next run.
\Oca thus guarantees
that reported invariants are side-effect-free modulo the test assertions, but a side
effect that no test exposes can still go undetected.

\subsubsection{Nested Invariant Invocation (C2.2)}
\todo{This section is hard to parse.  I think it would be improved by an example.}

Even a side-effect-free candidate can distort the invariants \Oca
reports, because evaluating it passes the program's own routines values
they would not otherwise see. Consider
\codeid{this\-.conversion\-Service\-.convert(source,} \codeid{sourceType,} \codeid{targetType)} \codeid{!=} \codeid{null},
proposed at \codeid{convert}'s exit:
evaluating it invokes the conversion service again,
which may dispatch to a different, instrumented converter and evaluate
that converter's own invariants against an execution that exists solely
because this (possibly incorrect!) candidate was being checked.

\Oca guarantees that only one invariant evaluates at a time on a given
thread: nested evaluation is suppressed, so one invariant's evaluation
cannot trigger another's.

\subsubsection{Nontermination (C2.3)}

Evaluating a candidate can also fail to terminate.
It
may call a method with an argument value that arises only
because of the evaluation itself{\,---\,}such as a much larger bound
than the method was designed for{\,---\,}and run
too long. %

\Oca monitors each run for signs that it has stalled.
On detection, it kills the run, permanently disables the candidate whose
probe was executing at that point, if any, and retries with
a larger time budget, repeating until a run finishes cleanly.

\section{Selecting a Prompt Strategy and Context (C1)}
\label{sec:config-selection}

We empirically evaluated prompt strategies and context configurations.
We organized this configuration-selection process around two research
questions:

\begin{itemize}
\item \textbf{RQ1 (Prompt Strategy):} Which prompt strategy is most
effective for generating invariants that hold during 
invariant evaluation?

\item \textbf{RQ2 (Context Configuration):} Which program-context
  configuration is most effective for generating invariants
   that hold %
under the selected prompt strategy?

\end{itemize}

\noindent We answer the two questions in sequence: RQ1 compares prompt strategies
under a fixed context configuration, and RQ2 compares context
configurations under the prompt strategy selected in RQ1. Testing every
pair of strategy and configuration would have multiplied the experimental
cost, so this design assumes the two factors are independent
(a threat to validity discussed in \cref{sec:limitations}).

For both RQs, we operationalized effectiveness as the number of
\emph{held invariants} produced by \Oca under the given configuration.
This is an objective and scalable metric: it can be measured without human
judgment. It does not measure invariant quality, however: a trivially true
invariant counts the same as a semantically meaningful one. Analogously to
statement coverage in testing, a configuration that yields
few held invariants is unlikely to be useful, but a high count alone does not
ensure useful invariants. We assess quality separately, through
expressiveness (\cref{sec:expressiveness-comparison}), false positives
(\cref{sec:false-positive-analysis}), and usefulness in property-based testing
(\cref{sec:usefulness}).

\subsection{Benchmark Projects}
\label{sec:config-benchmarks}

We searched open-source projects hosted on GitHub
using SEART's repository search infrastructure~\cite{seart},
applying the following criteria:
(1) the project is written in Java,
(2) it has at least 1{,}000 commits,
(3) it has at least 1{,}000 pull requests,
(4) it has at least 1{,}000 GitHub stars, and
(5) it has at least one commit in the three weeks before we
downloaded it.
\Oca targets Java programs, requiring criterion 1.
Criteria 2--5 favor mature, active, and widely used projects.
These criteria yielded 16 candidate projects.
We then excluded 3 repositories
whose available Javadocs were not predominantly written in English
(e.g., \codeid{alibaba/druid}).
Some context configurations in our study provided Javadocs to the LLM,
so the language of those Javadocs could itself affect
the generated invariants.
Including such projects
would therefore make it difficult to determine
whether differences between configurations were caused by the provided context
or by its language.
Next, we excluded 7 projects
for which the flaky-test detector iDFlakies~\cite{lam2019idflakies}
could not be run successfully.
Our experiments use iDFlakies to identify and remove
potentially flaky tests, which is important
because \Oca uses test failures to detect side effects.
After these filtering steps, six projects remained: Apollo~\cite{apollo},
Dubbo~\cite{dubbo}, Hudi~\cite{hudi}, libGDX~\cite{libgdx},
Netty~\cite{netty}, and Spring~\cite{spring}.
When a project contained multiple Java modules or subprojects,
we manually identified the primary module
containing the project's core functionality
and used that module in our evaluations.

\begin{table}[t]
\centering
\caption{Prompt strategies we evaluated. Each strategy other than Zero-Shot
  is the Zero-Shot template plus only the additions in its own row.
  Exact prompts are
  available in our replication package~\cite{OcaArtifact}.
}
\small
\begin{tabularx}{\columnwidth}{p{0.24\columnwidth} X}
\toprule
\textbf{Strategy \& \newline Prompting Technique} &
\textbf{Our Template} \tabularnewline
\midrule

Zero-Shot prompting \newline
Zero-shot &
The system message gives the task description and global rules
(output schema, allowed names, side-effect freedom, no tautologies).
The user message gives the program point, in-scope names, and program
context, and requests up to $K$ invariants as JSON. \tabularnewline\hline

Few-Shot prompting \newline
In-context learning &
Zero-Shot template + 1--2 worked examples (program point $\rightarrow$
expected JSON) before the target instance. \tabularnewline\hline

Chain-of-Thought \newline
Thought generation &
Zero-Shot template + an instruction to reason internally about types,
evaluability, and implied constraints before answering.
\tabularnewline\hline

Stepwise \newline
Decomposition &
Zero-Shot template + explicit sub-steps (unary, then relational, then
method-call invariants), combined into one JSON output. \tabularnewline\hline

Self-Refinement \newline
Self-criticism &
Zero-Shot template + an instruction to draft candidates, check each
against the rules, and return only the corrected set. \tabularnewline\hline

Multi-Sample \newline
Ensembling &
Zero-Shot template + an instruction to produce three internal drafts with
different focuses and keep only invariants that appear in at least two.
\tabularnewline

\bottomrule
\end{tabularx}
\label{tab:prompt-strategies}
\end{table}

\subsection{RQ1: Prompt Strategy}
\label{sec:prompt-strategy}

A single, unstructured instruction to propose an invariant tends to
lead to vague or trivially true output, since the model has no signal
about what kind of property is useful beyond ``something that holds''.
\Cref{tab:prompt-strategies} shows the text-based
prompting strategies that we evaluated instead, which we derived from the survey by
Schulhoff et al.~\cite{schulhoff2025prompt}. (Other surveys of prompt
engineering~\cite{huang2025prompt,sahoo2024systematic} are similar.)
We treated zero-shot prompting as the baseline strategy; every other
strategy extends it.

We instantiated each strategy with \textit{meta-prompting}~\cite{schulhoff2025prompt}:
we provided GPT-5 with a meta-prompt that combined a description of the
strategy with a task specification shared by all strategies and written
by the authors, and GPT-5 returned the strategy's prompt template.
The specification has three parts, which GPT-5 was instructed to copy
verbatim into the template.
\emph{Rules}, such as ``Each invariant must be a valid Java boolean
expression'', govern output format and well-formedness and go in the
system message.
\emph{Constraints}, such as ``Do not invent methods or fields that are
not available at this program point'', govern candidate content and go
in the user message, with the program point and context.
The third part, also in the user message, defines \codeid{result}, a
symbolic name for the method's return value along all paths.
Giving the specification to GPT-5, rather than appending it afterward,
lets each template build on it; e.g., Self-Refinement checks drafts
against the rules.

The complete rules, constraints, and generated prompt templates are
available in our replication package~\cite{OcaArtifact}.

\subsubsection{Experimental Results.}
We ran \Oca on each benchmark project (\cref{sec:config-benchmarks}),
varying only the prompt strategy while keeping all other factors fixed,
using the Local context configuration (\cref{sec:context-configuration}'s baseline).
As the LLM, we used \codeid{gpt-4.1-mini} via the OpenAI API,
with $K$ (the maximum number of invariants requested per query,
\cref{tab:prompt-strategies}) of 5.
\Cref{tab:prompt-results-approach} shows the number of held invariants that \Oca produces
for each combination of benchmark and prompt strategy.

\begin{table}[t]
\centering
\caption{Held invariants by prompt strategy, per project, using the Local
context configuration. Every
strategy was run on all six benchmark projects, so the final column
sums over all six and is directly comparable across rows.
Bold marks the best value in each column.}
\footnotesize
\begin{tabular}{lrrrrrrr}
\toprule
\textbf{Strategy} &
{\textbf{Apollo}} &
{\textbf{Dubbo}} &
{\textbf{Hudi}} &
{\textbf{libGDX}} &
{\textbf{Netty}} &
{\textbf{Spring}} &
{\textbf{Total}} \\
\midrule
Zero-Shot (baseline) & \textbf{1{,}515} & 8{,}119 & 316 & 711     & 1{,}045 & 5{,}937 & 17{,}643 \\
Few-Shot             & 1{,}396 & \textbf{9{,}017} & 257 & \textbf{1{,}673} & 909     & \textbf{6{,}021} & \textbf{19{,}273} \\
Chain-of-Thought     & 1{,}404 & 8{,}256 & 561 & 635     & 837     & 5{,}219 & 16{,}912 \\
Stepwise             & 1{,}378 & 8{,}328 & \textbf{580} & 672     & 453     & 4{,}973 & 16{,}384 \\
Self-Refinement      & 1{,}169 & 7{,}584 & 549 & 1{,}624 & \textbf{1{,}095} & 4{,}670 & 16{,}691 \\
Multi-Sample         & 638     & 3{,}972 & 271 & 978     & 440     & 2{,}355 & 8{,}654 \\
\bottomrule
\end{tabular}
\label{tab:prompt-results-approach}
\end{table}

Few-Shot produced the most held invariants overall (19{,}273);
most of the other strategies clustered around 17{,}000; and
Multi-Sample trailed the rest by a
wide margin (8{,}654). This ranking is not an artifact of aggregation:
Few-Shot was also the best strategy on three of the six
benchmark projects{\,---\,}Dubbo (9{,}017), libGDX
(1{,}673), and Spring (6{,}021){\,---\,}more projects than any other
strategy. So, we selected \textbf{Few-Shot} as the prompt strategy
for the rest of the paper.

\subsection{RQ2: Context Configuration}
\label{sec:context-configuration}

Independently of prompt strategy, the amount of source-level context
supplied affects what the model can reference when proposing an
invariant.
Unlike for prompt strategies, we could not find a survey of context
configurations. %
So, to select contexts, we conducted a literature review of the latest editions
of four top-tier software engineering conferences (FSE, ICSE, ASE, and ISSTA).

\newcommand{\taxn}[1]{\textcolor{black!55}{[#1]}}
\newcommand{\taxcat}[2]{\textbf{#1}~\taxn{#2}}
\newcommand{\taxsub}[4][9.25em]{\makebox[#1][l]{\textsf{#2}~\taxn{#3}}#4}
\newsavebox{\taxbox}
\sbox{\taxbox}{\begin{forest}
  for tree={font=\scriptsize, folder, grow'=0, folder indent=0.4em,
    inner xsep=1pt, inner ysep=0.55pt, s sep=0pt,
    edge={draw=black!40, line width=0.4pt}},
  desc/.style={no edge, font=\scriptsize\itshape}
  [{\textbf{Context Provided to LLM}~\taxn{181}}
    [{\taxcat{Local Code Context (LC)}{155}}
      [{Program code given directly to the model.}, desc]
      [{\taxsub{LC-SUBSRC}{73}{partial function code}}]
      [{\taxsub{LC-FUNC}{76}{whole function or method}}]
      [{\taxsub{LC-STRUCT}{43}{class, module, or file}}]
      [{\taxsub{LC-IR}{4}{AST, bytecode, or IR}}]]
    [{\taxcat{Structural Expansion Context (SE)}{42}}
      [{Context derived from program analysis.}, desc]
      [{\taxsub{SE-ANALYSIS}{31}{static analysis, types}}]
      [{\taxsub{SE-GRAPH}{14}{dependency graphs}}]]
    [{\taxcat{Documentation \& Retrieval (DOC/RET)}{112}}
      [{Non-code artifacts or retrieved knowledge.}, desc]
      [{\taxsub{DOC-NL}{56}{natural-language docs}}]
      [{\taxsub{DOC-SPEC}{40}{specs, APIs, standards}}]
      [{\taxsub{RET-CODE}{53}{code from repositories}}]
      [{\taxsub{RET-KB}{6}{curated knowledge bases}}]]
    [{\taxcat{Historical/Temporal Context (HIST)}{62}}
      [{Software evolution and development history.}, desc]]
    [{\taxcat{Semantic Context (SEM)}{69}}
      [{Abstractions capturing meaning or intent.}, desc]]
    [{\taxcat{Testing \& Execution Context (TEST)}{48}}
      [{Artifacts from program execution or testing.}, desc]
      [{\taxsub{TEST-EXEC}{32}{failures, traces, tests}}
        [{\taxsub[7.75em]{TEST-RENDER}{1}{rendered UI}}]]
      [{\taxsub{TEST-IO}{9}{input/output pairs}}]
      [{\taxsub{TEST-CODE}{18}{existing test suites}}]]
  ]
\end{forest}}
\begin{wrapfigure}{R}{\wd\taxbox}
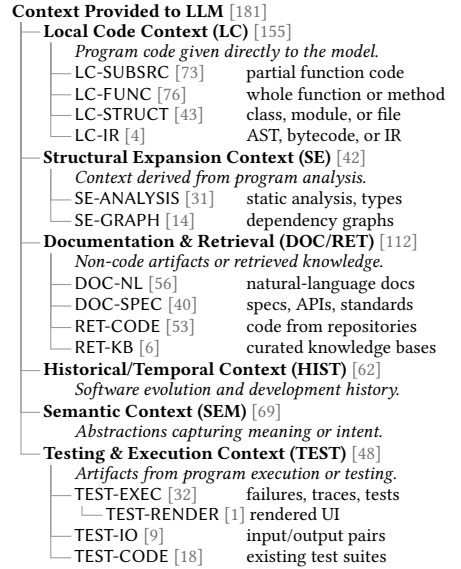

\usebox{\taxbox}
\caption{Taxonomy of context types provided to LLMs, derived via open coding
of prior work. Bracketed numbers count the 181 coded papers that use each
type or any of its subtypes; a paper can use several types, so parent counts
are unions rather than sums.}
\label{fig:context-taxonomy}
\vspace{-\baselineskip}
\end{wrapfigure}

\subsubsection{Literature Review Procedure.}
We first searched for
papers mentioning ``LLM'' or ``large language
model'' in their title or abstract, yielding 332 papers.
We manually examined these papers to identify those in which an LLM
performed a task on source code (e.g., program repair,
refactoring, code generation, or reasoning about program behavior),
resulting in 184 candidate papers.
We open-coded a random sample of these papers, starting with 20
(about 10\%) and adding five randomly selected papers at a time until
a batch of five produced no new category, which happened at 40 papers.
We examined the context supplied to the LLM in each sampled paper and
inductively grouped the observed types into the categorization in
\cref{fig:context-taxonomy}. We then applied these codes
to the remaining 144 papers, examining their methods and prompts where
available. One author performed the initial coding, and the other authors reviewed
the resulting codes; we resolved disagreements through live discussion
among the authors. %
The original categories covered the code-related context in every paper but one: a
frontend-repair study~\cite{yuan2025designrepair} supplied properties of the rendered interface.
We added the TEST-RENDER subcategory of TEST-EXEC
for this execution-derived context after the initial coding.
We could not assign a single study-level LLM-context code to three of the remaining papers,
because they compare several systems, describe no implemented system, or give no input to an LLM,
leaving 181 coded papers.
Our search results, screening decisions, per-paper codes, and
the justifications for them
are in our replication package~\cite{OcaArtifact}.

\subsubsection{Context Configurations Evaluated.}
We derived the context configurations from the six top-level
categories in \cref{fig:context-taxonomy}. %
We excluded \textbf{Historical/Temporal Context (HIST)} because
\Oca only takes source code as input, not version control history.
We also excluded \textbf{Semantic Context (SEM)} because the reviewed
papers derive it from hand-constructed abstractions (e.g., manually
written specifications) %
rather than mechanically extracted context; %
\Oca is fully automated.
For each of the four remaining categories, %
we selected one or
more representative context components that are (a) mechanically
extractable from the project %
and (b) plausibly informative for proposing an invariant at a given program point. Below, we
describe each by category, tagged with its taxonomy code(s).

We drew two components from the \textbf{Local Code Context (LC)} category as the baseline:
(1) the \textit{method body} (LC-FUNC): the source code of the method itself; and
(2) the \textit{scope} (LC-SUBSRC): a list of the in-scope variables and their types at the
program point.
Together, these give the model a concrete starting point, which helps prevent hallucinations.

\begin{table}[t]
\centering
\caption{Context configurations considered in our study.}
\small
\begin{tabular*}{\columnwidth}{@{}p{2.2cm}p{11.2cm}@{}}
\toprule
\textbf{Configuration} & \textbf{Composition} \\
\midrule

Local & Method body + in-scope variables (baseline). \\

Documented & Local + the method's own Javadoc. \\

Class-Aware & Local + enclosing class Javadoc, field names/types,
and method signatures (and Javadoc). \\

Type-Aware & Local + Class-Aware for all types referenced in the method's signature. \\

Usage-Aware & Local + call sites of the method with each
caller's Javadoc and implementation. \\

Example-Driven & Local + argument/return-value pairs observed from a
test run. \\

\bottomrule
\end{tabular*}
\label{tab:context-configurations}
\end{table}

We drew three components from the
\textbf{Structural Expansion Context (SE)} category{\,---\,}class documentation,
type documentation, and call-site context{\,---\,}all instantiating
SE-ANALYSIS: structural information about the program obtained by
statically resolving declarations and usages, rather than reasoning about
the method body in isolation.
The \textit{class documentation} (SE-ANALYSIS + DOC-NL, applied to the
enclosing class) consists of structured information about the class enclosing the
method: its Javadoc, the names and types of its fields, and the
signatures and Javadocs of its other methods. We treat this as an SE rather than DOC-NL component because,
although it does include some natural-language documentation,
it is assembled using an analysis.
The \textit{type documentation} (SE-ANALYSIS + DOC-NL, applied recursively to
referenced types) contains the same structured information{\,---\,}Javadoc, fields,
method signatures{\,---\,}but gathered recursively for the types referenced
in the method's signature (i.e., its parameter and return types). This
documentation enables the model to include an operation defined on a parameter's
type that is not visible from the method body or the enclosing class alone.
Finally, the \textit{call-site context} (SE-ANALYSIS, applied to call sites)
consists of call sites of the method elsewhere in the project, each paired with the
caller's own Javadoc and implementation. This shows the model how the
method is used in practice, which could suggest a constraint on
its inputs or outputs that is satisfied by every caller but not
stated in the method's own documentation.

We drew one component directly from the \textbf{Documentation \& Retrieval Context (DOC/RET)}
category{\,---\,}the \emph{method Javadoc} (DOC-NL). DOC-NL also appears combined with SE-ANALYSIS in two of the SE
components above, but this component is scoped to the method itself
rather than requiring resolution of other declarations.

Lastly, we drew one component from the \textbf{Testing \& Execution Context (TEST)}
category{\,---\,}\emph{I/O examples} (TEST-IO) observed from a test run.
These argument and return-value pairs were drawn from an index pre-built with
Daikon's Chicory~\cite{ErnstPGMPTX2007}. This gives the model concrete
evidence of what the method's inputs and outputs look like,
which a static reading of the source may not provide.

We combined these components into named \emph{context configurations},
which each supply a different mix of context to the prompt. Every
configuration includes the method body and scope (LC) as a baseline,
which we call \textit{Local}. Each remaining configuration adds exactly one non-LC component to Local:
\textit{Documented} adds the method Javadoc, \textit{Class-Aware} adds
class documentation, \textit{Type-Aware} adds type documentation,
\textit{Usage-Aware} adds call-site context, and \textit{Example-Driven}
adds I/O examples. \Cref{tab:context-configurations} summarizes the resulting
configurations.

\subsubsection{Experimental Results.}
All configurations were evaluated with the Few-Shot prompt
strategy selected in \cref{sec:prompt-strategy}.
\Cref{tab:context-results-approach} shows the results.
Class-Aware produced the most held invariants overall (20{,}518).
Based on these results, \Oca uses the \textbf{Few-Shot} prompt strategy with the
\textbf{Class-Aware} context configuration in all subsequent experiments in this paper.

\begin{table}[t]
\centering
\caption{Held invariants by context configuration, per project,
under the selected Few-Shot prompt strategy. \textit{Local} reuses the
Few-Shot row from \cref{tab:prompt-results-approach}, since it is the
same underlying configuration. Bold marks the best value in each column.
}
\small
\begin{tabular}{lrrrrrrr}
\toprule
\textbf{Configuration} &
{\textbf{Apollo}} &
{\textbf{Dubbo}} &
{\textbf{Hudi}} &
{\textbf{libGDX}} &
{\textbf{Netty}} &
{\textbf{Spring}} &
{\textbf{Total}} \\
\midrule
Local (baseline) & 1{,}396 & 9{,}017 & 257 & 1{,}673 & 909     & 6{,}021 & 19{,}273 \\
Usage-Aware      & \textbf{1{,}554} & \textbf{9{,}414} & 255 & 941     & 824     & 6{,}173 & 19{,}161 \\
Class-Aware      & 1{,}543 & 9{,}384 & 332 & \textbf{1{,}819} & \textbf{1{,}021} & \textbf{6{,}419} & \textbf{20{,}518} \\
Documented       & 1{,}401 & 9{,}045 & 262 & 577     & 880     & 6{,}230 & 18{,}395 \\
Type-Aware       & 1{,}396 & 9{,}067 & 257 & 751     & 908     & 6{,}038 & 18{,}417 \\
Example-Driven   & 1{,}548 & 8{,}868 & \textbf{333} & 1{,}046 & 872     & 6{,}137 & 18{,}804 \\
\bottomrule
\end{tabular}
\label{tab:context-results-approach}
\end{table}

To understand why Class-Aware produced more held invariants, we manually
inspected nine randomly sampled methods for which every configuration
other than Local produced at least one held invariant, plus
\CollectionToObjectConverter{}\codeid{.convert} from our running example
(\cref{sec:running-example}). In most of these methods, Class-Aware's
invariants were more semantically meaningful, captured richer
relationships among program values or object state, or avoided trivial
and observation-specific properties that other configurations produced.
For example, Class-Aware was the only configuration to recover the key
relationship in \codeid{convert}: when the source type is assignable to
the target type, the result is the source object itself. A plausible
explanation is that class-level context describes the surrounding
abstraction, %
letting the model connect the method to relationships not explicit in
its body. The other configurations showed characteristic biases:
Example-Driven sometimes anchored on salient values in the supplied
executions, such as nullness, numeric ranges, or particular run-time
states, and Usage-Aware sometimes emphasized relationships specific to
how the observed callers use the method. Our inspection thus complements
the quantitative result: class-level context tends to yield more
informative candidates, not only more of them.

\section{Filtering and Evaluating Candidates}
\label{sec:checking}

This section describes how \Oca
statically filters and compiles candidate invariants proposed by the LLM to discard malformed ones (C1),
then evaluates them safely during test execution (C2).

\subsection{Filtering and Compiling Candidates (C1)}
\label{sec:filtering}

\subsubsection{Static Filtering}
\label{sec:static_filtering}
Before injecting a candidate, \Oca checks it against a set of
rejection criteria. A candidate is rejected
if it exceeds a length bound
intended to catch degenerate or runaway generations;
if it is a syntactic
tautology, such as a disjunction of an expression and its
negation;
if it uses a construct that
may introduce further side effects or nontermination on its
own (lambdas, streams, object construction, \codeid{throw},
\codeid{System.exit});
if it mentions no variable in scope at that program
point;
if it performs a
null check against a variable of primitive type; %
if it references an identifier that is not in scope; %
or
if it does not parse as a valid Java expression.

\subsubsection{Compilation and Error Attribution}
\label{sec:compilation}
After injecting candidate invariants into the program, \Oca
runs the compiler and parses any resulting compiler errors.
\Oca wraps each injected probe in begin/end marker comments, so each
delimited region contains the probe's code, including the candidate's
identifier (\cref{sec:probe-structure}). When the compiler reports an
error inside such a region, \Oca removes the region, records the
candidate as failing to compile, and retries.
If errors remain after a configurable number of retries, \Oca
restores each file that still fails to compile to its uninstrumented
state, guaranteeing that this process terminates with a successful compilation.

\subsection{Evaluating Candidates Safely (C2.1--C2.3)}

\Cref{sec:overview-safety} introduced challenges C2.1--C2.3, which arise
from evaluating candidates.

\subsubsection{Probe Structure}
\label{sec:probe-structure}

\begin{wrapfigure}{R}{0.36\textwidth}
\begin{lstlisting}[basicstyle=\ttfamily\scriptsize, frame=single, escapeinside={(*}{*)}]
disabled: global map, id to boolean,
          initially false
evaluating: thread-local flag,
            initially false
(*\makebox[0pt][l]{\rule[0.5ex]{\linewidth}{0.4pt}}*)
probe(id, invariant):
  if disabled[id] or evaluating:
    return
  evaluating := true
  try:
    ok := invariant
  catch any exception:
    ok := false
  finally:
    evaluating := false
  if not ok:
    record id as falsified
    disabled[id] := true
\end{lstlisting}
\caption{The bottom part is the structure of the probe injected for a candidate \codeid{invariant} with
identifier \codeid{id}.
For clarity, the figure omits bookkeeping for hang detection
(\cref{sec:hangs}) and for carrying outcomes across processes.}
\label{fig:probe-pseudocode}
\vspace{-\baselineskip}
\end{wrapfigure}

Each candidate is
injected as one or more \emph{probes}; \cref{fig:probe-pseudocode} shows a
probe's structure in the abstract. \Oca represents each candidate with a unique identifier
shared by its probes.
Once a candidate is falsified, its probes stop evaluating it for the
rest of the process. \Oca also records falsified candidates, and those
that hang (\cref{sec:hangs}) or cause a test failure
(\cref{sec:side-effects}), so that later processes skip them.
Within one run, whenever execution reaches a candidate's
program point, its expression is evaluated unless the candidate has been
disabled or another invariant is already evaluating on the same thread.
A method-entry probe is
placed at the start of the method body; a method-exit probe is
placed immediately before each return statement, or at the point of
fall-through for a void method. 

\subsubsection{Side Effects}
\label{sec:side-effects}

\Oca uses newly failing tests as a proxy for side effects
from candidate invariants (C2.1). We would prefer that \Oca statically prove
that candidates are side-effect free, but the available tools are so conservative
that they reject most candidates in practice, leading \Oca to discard too many
useful candidates~\cite{helm2018unified}. \Oca's approach underapproximates side effects, since one that
no test exercises goes undetected. In exchange, it catches side
effects that disrupt the behavior developers actively validate.

When a test fails on an instrumented run, \Oca applies the \textit{ddmin}
delta-debugging algorithm~\cite{zeller2002simplifying} to the set of candidates
executed during that run, reducing it to a 1-minimal set of culprits.
\Oca orders the candidates so that those executed closest to the failure
come first, on the premise that they are more likely responsible; this
biases the search toward finding the culprits early. Once a culprit set is
isolated, \Oca disables
it and reruns the test suite to confirm that the
failure no longer reproduces, which guards against coincidental reductions.

\subsubsection{Nested Evaluation}
\label{sec:nesting}

Evaluating a candidate can call a method that has its own injected probe,
so the outer evaluation triggers the inner probe's evaluation against
inputs that exist only because the outer candidate is being checked. This
nesting leads to two challenges: the inner invariant
may be falsified incorrectly (C2.2), and if the method
containing the inner probe transitively calls back into the method
containing the outer probe, the two probes can trigger each other
indefinitely, hanging the run (C2.3).
The \codeid{evaluating} check in \cref{fig:probe-pseudocode} closes off both
nested cases: this thread-local flag is set while a
probe evaluates its invariant, so a probe reached during another
invariant's evaluation returns without evaluating its own.

\subsubsection{Detecting and Recovering from Hangs}
\label{sec:hangs}

Non-termination is still possible without nesting: a candidate can call
a method that blocks indefinitely without involving any other probe.
\Oca kills a run if it produces no output on standard output or standard
error for longer than a silence timeout, or if it exceeds its time budget.
If a candidate was being evaluated at that moment, \Oca disables it.
Then, \Oca reruns the program from the start with a longer time budget,
doubled from the previous attempt up to a fixed cap, so that a few
retries suffice for a slow but legitimate run.
If a run at the cap is killed while no candidate is being evaluated,
\Oca by default disables the most recently evaluated candidate and reruns.
Alternatively, \Oca can be configured to disable every candidate evaluated
before the stall, which prevents the same stall from recurring if the
uninstrumented tests finish within the cap, at the cost of not checking
those candidates again. It can also be configured to stop and report
the invariants held so far, leaving the user to diagnose the stall.

\section{Evaluation}
\label{sec:evaluation}

Our evaluation compares \Oca to Daikon in four ways:
\begin{itemize}
  \item \textbf{RQ3 (Expressiveness).} Does \Oca infer
    invariants that Daikon structurally cannot?

  \item \textbf{RQ4 (False-Positive Analysis).} Compared to Daikon, how many
    ``invariants'' reported by \Oca do not actually hold on all possible
    executions?

  \item \textbf{RQ5 (Usefulness).} How do the invariants inferred by \Oca
    compare to those inferred by Daikon when used for a downstream task,
    property-based testing?

  \item \textbf{RQ6 (Performance).} How does the cost of generating
    invariants with \Oca compare to the cost of generating invariants
    with Daikon?
\end{itemize}

\noindent Unless otherwise noted, \Oca
uses the prompt strategy and context configuration selected
in \cref{sec:config-selection} (\textbf{Few-Shot} prompting with \textbf{Class-Aware} context)
and Daikon uses its default settings
throughout.
All
experiments in this evaluation, for both Daikon and \Oca, were run
on a high-performance computing (HPC) cluster, on nodes with AMD EPYC 7753 CPUs and 128\,GB of RAM.
\Oca uses \codeid{gpt-4.1-mini} via the OpenAI API, with $K = 5$,
as in \cref{sec:config-selection}.

\subsection{Subjects and Datasets}
\label{sec:evaluation-subjects}

We drew on two sets of subject programs in this evaluation.
The first is the set of six benchmark projects from %
\cref{sec:config-benchmarks}
(Apollo,
Dubbo,
Hudi,
libGDX,
Netty,
and Spring).
Since this benchmark was also used to select \Oca's prompt strategy
and context configuration, there is a risk of overfitting, which motivates
the inclusion of a second dataset.
The second dataset is \emph{Defects4J}~\cite{JustJE2014},
a widely used benchmark suite of real-world Java programs with
curated buggy and fixed versions and high-quality test suites.
For each subject program,
Defects4J provides paired buggy and fixed revisions,
enabling controlled comparisons across program versions.

Defects4J contains 17 projects. We excluded Chart because checking it out
requires Subversion, which was unavailable on our cluster. We then piloted
the usefulness experiment in \cref{sec:usefulness} on the latest bugs of the remaining 16 projects. A
project was included only if every step completed successfully for both
tools within the cluster's 72-hour walltime limit.
Daikon's Chicory frontend exceeded this limit while %
collecting traces on 10 projects, and crashed on another (Jsoup).
We therefore experimented with only five projects: Cli, Codec, Collections, Gson, and Math.

\subsection{RQ3: Expressiveness}
\label{sec:expressiveness-comparison}

The subject projects in this experiment are the six benchmarks from
\cref{sec:config-selection} and the five Defects4J projects;
we analyze and report the two sets separately.
For every subject project,
we ran both Daikon and \Oca using the same project version and
test suite,
inferring invariants at all method entry and exit program points.
The same executions were used for both tools wherever possible,
so that differences in the resulting invariant sets reflect
differences between the invariant-detection approaches rather than
differences in exercised program behavior.

The independent variable is the invariant-inference tool---Daikon or \Oca.
The dependent variable is invariant expressiveness:

\begin{definition}[Expressiveness]
We define an inferred invariant as \emph{expressive} if it
(1) involves project-specific operations or
(2) relates more than three variables.
\end{definition}

\begin{wraptable}{R}{0.54\textwidth}
\centering
\caption{Expressiveness results for the benchmark projects used in the
prompt and context studies and for the fixed Defects4J projects.
Each $V_{>3}$ and PSPM entry gives the count of held \Oca invariants of that
kind and, in parentheses, its share of $N_O$; percentages in the
\textbf{Total} rows are macro-averages.
\todo{This table should be on the same page as the start of the RQ3 results paragraph before we submit.}
}
\label{tab:invariant-stats}
\small
\setlength{\tabcolsep}{4pt}
\begin{tabular}{lrrrr}
\toprule
\textbf{Project} & $\boldsymbol{N_D}$ & $\boldsymbol{N_O}$ &
$\boldsymbol{V_{>3}}$ & \textbf{PSPM} \\
\midrule
\multicolumn{5}{l}{\textit{Benchmark projects (prompt/context study)}} \\
Apollo & 59{,}316 & 1{,}543 & 53 (3\%) & 237 (15\%) \\
Dubbo$^\dagger$ & 19{,}730 & 9{,}384 & 378 (4\%) & 4{,}091 (44\%) \\
Netty$^\dagger$ & 75{,}756 & 1{,}021 & 48 (5\%) & 215 (21\%) \\
Hudi & 312{,}607 & 332 & 2 (1\%) & 6 (\phantom{0}2\%) \\
libGDX$^\dagger$ & 21{,}101 & 1{,}819 & 101 (6\%) & 329 (18\%) \\
Spring & 9{,}590 & 6{,}419 & 194 (3\%) & 2{,}569 (40\%) \\
\midrule
\textbf{Total} & 498{,}100 & 20{,}518 & 776 (4\%) & 7{,}447 (23\%) \\
\midrule
\multicolumn{5}{l}{\textit{Defects4J projects (fixed versions)}} \\
Cli & 6{,}389 & 1{,}929 & 87 (5\%) & 513 (27\%) \\
Codec & 45{,}147 & 4{,}049 & 257 (6\%) & 1{,}231 (30\%) \\
Collections & 122{,}046 & 18{,}549 & 1{,}071 (6\%) & 7{,}975 (43\%) \\
Gson & 418{,}757 & 3{,}426 & 231 (7\%) & 1{,}021 (30\%) \\
Math & 18{,}789 & 5{,}522 & 337 (6\%) & 1{,}990 (36\%) \\
\midrule
\textbf{Total} & 611{,}128 & 33{,}475 & 1{,}983 (6\%) & 12{,}730 (33\%) \\
\bottomrule
\end{tabular}

\vspace{2pt}
\parbox{\linewidth}{\footnotesize
$^\dagger$Daikon ran on a sampled trace (\codeid{--sample-start=5})
because tracing did not finish within the 72-hour walltime
limit (Dubbo, libGDX) or failed with an exception (Netty).}
\end{wraptable}

\noindent This definition intentionally \emph{underestimates \Oca's full
expressive capability},
since it excludes other forms of semantic richness
that may also lie outside Daikon's fixed template library.
We adopt this restricted definition to make the comparison with
Daikon objective and unambiguous.
Both criteria correspond to expressiveness boundaries that Daikon
cannot cross by design:
Daikon's invariant grammar does not support project-specific method
calls and limits relations to a small,
fixed number of variables.
By focusing on these dimensions,
we avoid subjective judgments about invariant quality and ensure
that any measured expressiveness advantage reflects capabilities
fundamentally unavailable to Daikon.

For each project,
we recorded four quantities:
$N_D$,
the total number of held invariants reported by Daikon;
$N_O$,
the total number of held invariants reported by \Oca;
$V_{>3}$,
the number of held \Oca invariants that relate more than three
variables;
and $\mathrm{PSPM}$,
the number of held \Oca invariants that contain at least one
project-specific method call,
which we used to operationalize the project-specific-operation
criterion above.
From these,
we computed two expressiveness metrics for \Oca:
the \emph{Var$>$3 percentage},
$V_{>3}/N_O$,
and the \emph{PSPM percentage},
$\mathrm{PSPM}/N_O$.
We report $N_D$ and $N_O$ alongside these percentages for context, not
as a measure of relative quality, because the two counts are not
comparable. \Oca's count scales with $K$, the maximum number of
invariants requested per query, which we set to 5
(\cref{sec:config-selection}); raising $K$ yields more held invariants.
Daikon's count reflects its large template grammar, which instantiates
many low-information invariants, such as the only invariant it reports
for our running example: that \codeid{this.conversionService} is not
side-effected (\cref{sec:overview}).

For three of the configuration-selection benchmark projects, Daikon
could not produce invariants from a full trace. For Dubbo and libGDX,
tracing with Chicory did not finish within the 72-hour maximum walltime
of our institution's HPC cluster; for Netty,
Daikon terminated with an exception while processing the full trace.
For these projects, we reran
Chicory with \codeid{--sample-start=5}, which records every execution
of a program point until it has executed five times and samples later
executions, and ran Daikon on the resulting trace. Their $N_D$ values
therefore come from sampled traces, which can change the set of
invariants Daikon reports in both directions: an invariant that a
skipped execution would falsify can survive, and an invariant can go
unreported because too few observations remain for Daikon to justify
it statistically.

\subsubsection{RQ3 Results}

\Cref{tab:invariant-stats} gives the results.
Overall, the results support an expressiveness advantage for \Oca.
By design, none of Daikon's invariants relate more than three variables
or call a project-specific method; \Oca has no such limitations.
In fact, \Oca produced such
invariants on every project: on the benchmark projects, a macro-average
of 4\% of its held invariants related more than three variables and
23\% called a project-specific method; on Defects4J, the corresponding
figures were 6\% and 33\% (5\% and 28\% across all 11 projects). These percentages underestimate \Oca's expressiveness,
since the definition deliberately counts only these two forms. \Oca's
advantage is the kinds of properties it can express, not the
number of invariants, which is not directly comparable between the tools.

\subsection{RQ4: False-Positive Analysis}
\label{sec:false-positive-analysis}

Both Daikon and \Oca are dynamic invariant detectors:
an invariant that they report ``holds'' was true on the observed executions,
but it is not necessarily a semantic property of the program.
An incorrect candidate may therefore be reported as held simply
because the available test executions do not exercise an input that
falsifies it.
Since Daikon and \Oca propose invariants in different ways,
one tool could introduce more ``false-positive'' invariants
that are not true on all possible program executions.
\looseness=-1

To test whether one tool does, we manually analyzed a stratified sample of 100 held invariants,
50 from each tool,
drawn from the outputs of Daikon and \Oca on the benchmark
projects described in \cref{sec:config-benchmarks}.
The 100 invariants come from 19 distinct methods across all the
benchmark projects.
We stratified the sample across projects and methods,
including methods exercised with different numbers of inputs,
to avoid concentrating the analysis on a small number of projects,
methods,
or highly exercised program points.

\begin{wraptable}{R}{0.46\textwidth}
\centering
\caption{False-positive analysis of the manually inspected sample
(RQ4).}
\label{tab:false-positives}
\small
\begin{tabular}{lrrrr}
\toprule
\textbf{Tool} & \textbf{False pos.} & \textbf{Valid} &
\textbf{Total} & \textbf{Rate} \\
\midrule
\Oca   & 9  & 41 & 50 & 18\% \\
Daikon & 38 & 12 & 50 & 76\% \\
\bottomrule
\end{tabular}
\end{wraptable}

To reduce bias,
one author prepared the sampled invariants for
blinded review by removing all information identifying which tool produced each.
Because Daikon reports invariants in its own textual notation,
this author rewrote the sampled Daikon invariants as semantically
equivalent Java expressions so that their representation was
indistinguishable from that of the \Oca invariants.
The resulting blinded sample was independently inspected by two
other authors.
For each invariant,
the reviewers examined the corresponding program point and source
code and determined whether the invariant was \emph{semantically
valid}:
that is,
whether it followed from the program's semantics and the expressed
programmer intent at that program
point rather than merely holding for the executions observed by the
dynamic analysis.
An invariant was classified as a \emph{false positive} when it was
reported as held by the tool but was not semantically valid and had
survived only because the available executions did not expose a
counterexample.
Disagreements between the two reviewers were resolved by the
judgment of a third author.

We report,
for each tool,
the number and percentage of sampled invariants classified as false
positives.
We additionally report the number and percentage of classifications
on which the two reviewers agreed,
and their inter-rater agreement according to Cohen's $\kappa$,
both computed on the reviewers' independent classifications,
before disagreements were resolved.

\subsubsection{RQ4 Results}

\Cref{tab:false-positives} shows the results.
Of the 50 sampled Daikon invariants,
38 (76\%) were classified as false positives,
compared with 9 of 50 (18\%) for \Oca.
The reviewers agreed on 89 of the 100 classifications
(89\% agreement) and disagreed on 11 (11\%),
which the third author resolved.
Their inter-rater agreement was substantial,
with a Cohen's $\kappa$ of 0.780.
Overall,
\Oca had the lower observed false-positive rate in this manually
inspected sample.

\subsection{RQ5: Usefulness via Simulated Property-Based Testing}
\label{sec:usefulness}

We evaluated invariant usefulness by simulating a realistic
debugging workflow in which invariants are inferred \emph{before} a
bug-revealing input is known and are then evaluated once such an
input becomes available. The goal is to infer an invariant that the original programmers
\emph{intended} to always be true but that only the bug-revealing input
falsifies (because of the bug).
For this experiment, we used buggy versions of the Defects4J benchmarks.
While working on this experiment, we resolved five issues that caused Daikon to fail, but additional
randomly selected pilot bugs revealed further issues that required manual
intervention. Daikon also required up to 1\,TB of traces per
bug, which limited how many bugs could run concurrently.
Given these execution and storage limitations and the manual effort needed to resolve failures, we limited our evaluation to the
latest five bugs of each included project, 25 bugs in total.
For each bug, we applied the same procedure to Daikon and \Oca:

\begin{enumerate}
  \item \textbf{Infer candidate invariants.} Run the buggy version's test suite
  without the bug-revealing tests and retain the held
  candidate invariants. %

  \item \textbf{Check whether the bug-revealing tests falsify any invariants.} Run the bug-revealing tests and
  identify the candidates falsified by them. %

  \item \textbf{Check the fixed version.} Check that those candidates are not falsified anywhere in the
  fixed version's full test suite.
\end{enumerate}

A bug is \emph{exposed} if at least one candidate satisfies all of these
requirements. We attribute each invariant check to the test that triggered it,
and neither tool infers new candidates after the first step.
Excluding the bug-revealing tests in the first step models a developer
who has not yet identified a failing test, and it prevents bug-triggering
inputs from leaking into inference.

\begin{wraptable}{R}{0.42\textwidth}
\centering
\caption{Bugs exposed per project
  (RQ5).}
\label{tab:usefulness-results}
\small
\begin{tabular}{lrrr}
\toprule
\textbf{Project} & \textbf{Bugs} & \textbf{\Oca} & \textbf{Daikon} \\
\midrule
Cli         & 5 & 0 & 0 \\
Codec       & 5 & 2 & 1 \\
Collections & 5 & 5 & 1 \\
Gson        & 5 & 1 & 0 \\
Math        & 5 & 1 & 2 \\
\midrule
\textbf{Total} & \textbf{25} & \textbf{9 (36\%)} & \textbf{4 (16\%)} \\
\bottomrule
\end{tabular}
\end{wraptable}

This evaluation design intentionally factors input generation out of
the experiment.
Although property-based testing typically relies on automated input
generation (e.g.,
random or coverage-guided fuzzing),
Defects4J already provides inputs that are known to expose each bug.
By reusing these inputs,
we isolated and compared the usefulness of the inferred invariants
themselves,
independent of the effectiveness of any particular input-generation
technique.
Both Daikon and \Oca were evaluated on identical bug-revealing
inputs,
ensuring a fair and controlled comparison.

We report the number of bugs each tool exposes per project, and its
\textbf{Bug Exposure Rate} (exposed bugs divided by evaluated bugs) in aggregate.

\subsubsection{RQ5 Results}

\Cref{tab:usefulness-results} reports the bugs each tool exposed.
\Oca exposed 9 of the 25 bugs (36\%), and Daikon exposed 4 (16\%).
Overall, the results support a substantial usefulness advantage for
\Oca: it exposed more than twice as many bugs as Daikon
(9 versus 4), and it exposed at least as many bugs as Daikon on every
project except Math. Because the comparison covers 25 bugs from five projects, the
size of the advantage should be interpreted with care
(\cref{sec:limitations}).

\paragraph{Illustrative example: Math-105.}

Math-105 illustrates a bug exposed by both tools.
The affected method,
\codeid{SimpleRegression.getSumSquaredErrors},
returns the sum of squared errors (SSE) of a regression model.
Its documentation specifies that negative results caused by rounding
are returned as zero, and that \codeid{NaN} is returned when
insufficient observations have been added to estimate the model.
With the bug-revealing tests excluded, both tools inferred
\codeid{result >= 0.0} at the method's exit.
\Oca also inferred the following entry property:

{\lstset{keywordstyle=[3]{\normalcolor}}
\begin{lstlisting}[frame=tb,aboveskip=3mm,belowskip=3mm]
Double.isNaN(getSumSquaredErrors()) || getSumSquaredErrors() >= 0
\end{lstlisting}}

\noindent
This entry property calls \codeid{getSumSquaredErrors} itself;
because the method is side-effect-free, this constrains the value
it is about to return.
\Oca's suppression of nested invariant evaluation
(\cref{sec:nesting}) prevents these calls from recursively
evaluating the same invariant.

The buggy implementation computes
\codeid{sumYY - sumXY * sumXY / sumXX}.
Although an SSE is nonnegative in exact arithmetic,
floating-point cancellation produces approximately
$-7.1 \times 10^{-15}$ for the bug-revealing input,
contradicting the method's documentation and falsifying
all three candidates.
The fix wraps the expression in \codeid{Math.max(0d, ...)},
returning zero for the same input.
All three candidates satisfied the fixed-version
check of our procedure, so both tools expose the bug.
\Oca additionally inferred a property that explicitly permits
the documented \codeid{NaN} outcome:
under Java's expression semantics,
the two exit properties reject \codeid{NaN},
whereas the entry property accepts it.
This property combines a project-specific method call with a
disjunction, which template-free inference makes possible,
and so accounts for both a nonnegative SSE and the documented
outcome when the model cannot yet be estimated.

\subsection{RQ6: Performance}
\label{sec:performance}

Unlike Daikon,
\Oca uses a new and relatively expensive component:
the LLM that proposes candidate invariants.
But
it evaluates far fewer candidate invariants
and records no execution traces to disk.
So, it is not obvious which of the two systems is more expensive.

 \begin{wraptable}{R}{0.485\textwidth}
\centering
\caption{Time and disk usage per project, on the same runs used to answer RQ3.}
\label{tab:performance-results}
\small
\setlength{\tabcolsep}{4pt}
\begin{tabular}{lrrrr}
\toprule
& \multicolumn{2}{c}{\textbf{Time [min]}} & \multicolumn{2}{c}{\textbf{Disk [GB]}} \\
\cmidrule(lr){2-3}\cmidrule(lr){4-5}
\textbf{Project} & \textbf{Daikon} & \textbf{\Oca} & \textbf{Daikon} & \textbf{\Oca} \\
\midrule
\multicolumn{5}{l}{\textit{Benchmark projects (prompt/context study)}} \\
Apollo & 8.1 & 14.4 & 2.56 & 0.38 \\
Dubbo$^\dagger$ & 23.9 & 129.9 & 1.34 & 0.78 \\
\quad\textit{full (failed)} & -- & -- & 280.00$^\ast$ & -- \\
Netty$^\dagger$ & 1{,}549.7 & 60.0 & 0.76 & 0.73 \\
\quad\textit{full (failed)} & -- & -- & 2{,}300.00 & -- \\
Hudi & 27.1 & 11.3 & 683.30 & 19.00 \\
libGDX$^\dagger$ & 350.6 & 83.0 & 90.32 & 2.81 \\
\quad\textit{full (failed)} & -- & -- & 1{,}900.00$^\ast$ & -- \\
Spring & 2.2 & 90.8 & 2.21 & 2.62 \\
\midrule
\textbf{Total} & 1{,}961.6 & 389.4 & 780.49 & 26.32 \\
\midrule
\multicolumn{5}{l}{\textit{Defects4J projects (fixed versions)}} \\
Cli & 10.0 & 9.1 & 0.60 & 0.02 \\
Codec & 570.0 & 14.5 & 14.20 & 0.03 \\
Collections & 240.0 & 101.1 & 26.30 & 0.13 \\
Gson & 20.0 & 10.3 & 3.10 & 0.03 \\
Math & 15.0 & 41.2 & 2.40 & 0.09 \\
\midrule
\textbf{Total} & 855.0 & 176.2 & 46.60 & 0.30 \\
\bottomrule
\end{tabular}

\vspace{2pt}
\parbox{\linewidth}{\footnotesize
$^\dagger$Daikon ran on a sampled trace; the \emph{full (failed)} row
gives the size of the full-trace attempt, excluded from the
totals.\\
$^\ast$Partial trace at the 72-hour limit.}
\vspace{-\baselineskip}
\end{wraptable}

We reused the Daikon and \Oca runs performed for RQ3
(\cref{sec:expressiveness-comparison}) rather than making dedicated
runs, so that the cost comparison reflects the exact executions
whose expressiveness results are reported there. For each project and
each tool, we recorded the wall-clock time and disk usage of the
invariant-inference run, computed differently for each tool to match
what each actually produces.
For \textbf{Daikon}, \emph{time} is the sum of the wall-clock time spent
running Chicory, Daikon's instrumentation and trace-collection tool, and
the wall-clock time spent running Daikon itself to infer invariants from
Chicory's traces. \emph{Disk usage} is the total size of Chicory's dtrace files,
together with the log files in which Daikon reports inferred invariants.
For \textbf{\Oca}, \emph{time} is the total wall-clock time for a
single end-to-end run of the tool on a project, including the LLM
querying phase (remote calls to the OpenAI API), filtering, compilation, and test-suite execution.
\emph{Disk usage} is the size of the copied, instrumented version of
the project plus the output and %
the log files \Oca produces.

\subsubsection{RQ6 Results}

\Cref{tab:performance-results} shows the results.
The full-trace attempts of the three $^\dagger$ projects cost far more before
failing: Netty's took 59.8 hours to collect and occupied
2.3\,TB before Daikon failed on it after 10.5 hours; libGDX's Chicory run
was stopped at the 72-hour walltime limit with 1.9\,TB of partial trace,
on which Daikon failed after 23.9 hours; and Dubbo's Chicory run likewise
did not complete, leaving a 280\,GB partial trace on which Daikon failed
after 69.1 hours.
\Oca also incurs a cost that Daikon does not: LLM token usage, since it
sends a prompt for each program point. Across all 11 projects, it used
about 107 million tokens, costing about \$55.

Overall, the results support a performance advantage for \Oca.
It was faster than Daikon on both subject sets (389.4 versus
1{,}961.6 minutes on our benchmarks and 176.2 versus 855.0 on
Defects4J) and used less disk (26.32 versus 780.49\,GB and 0.30 versus
46.60\,GB), with one exception (Spring). Time was mixed per project:
Daikon was faster on Apollo, Dubbo, Spring, and Math, while \Oca was
much faster where Daikon's runs were long, such as Netty, libGDX, and
Codec. \Oca's run time is also more predictable:
it scales roughly linearly with the number of program
points ($r = 0.90$),
whereas Daikon's run time is uncorrelated
with program-point count ($r = -0.01$).

\section{Limitations and Threats to Validity}
\label{sec:limitations}

\paragraph{Internal Validity}
\Oca only guarantees that a held invariant is side-effect-free
modulo the project's test suite (\cref{sec:side-effects}).
Similarly, our flaky-test filtering (\cref{sec:config-benchmarks}) may miss
some flaky tests, so a few test failures we
attribute to an injected candidate could instead reflect residual
flakiness.
Our prompt-strategy and context-configuration selection
(\cref{sec:config-selection}) used
a single LLM (\codeid{gpt-4.1-mini}); results could differ for
other models.
This selection also assumed that the two
factors are independent; an untested combination could outperform
the one we selected.

\paragraph{External Validity}
We evaluated only Java programs, so our findings may not generalize to other languages.
Our selection criteria favored mature,
popular, and active GitHub repositories (\cref{sec:config-benchmarks}), so the six benchmark projects
may not represent smaller or less active projects. We selected the
prompt strategy and context configuration on these same six projects
(\cref{sec:config-selection}), so results on them do not independently
confirm that this configuration is optimal; hence we also evaluate on Defects4J
(\cref{sec:evaluation-subjects}).
Our literature review (\cref{sec:context-configuration}) covered one
edition of each of four venues and searched titles
and abstracts, so it may miss papers that use an LLM without saying so
there, and its prevalence counts reflect recent work at these venues,
not the whole field.

\paragraph{Construct Validity}
For three benchmark projects, Daikon's invariant counts come from traces
sampled with Chicory's \codeid{--sample-start} option, because full
runs exceeded our cluster's 72-hour walltime
limit on two projects and crashed Daikon on the third
(\cref{sec:expressiveness-comparison}). Sampling can change which
invariants Daikon reports, so these counts are not directly comparable
to those from full traces.
Our expressiveness metric (\cref{sec:expressiveness-comparison})
counts an invariant as expressive only if it relates more than three
variables or invokes a project-specific method; by design, this
underestimates \Oca's expressiveness, since it ignores
other forms of semantic richness. Our false-positive analysis
(\cref{sec:false-positive-analysis}) relies on manual, blinded
judgments of semantic validity by two reviewers; although we report
Cohen's $\kappa$, such judgments remain subjective.
One author coded the context taxonomy (\cref{fig:context-taxonomy});
the others reviewed the codes and resolved disagreements by discussion, but we did not measure inter-rater agreement,
so the codes may reflect one coder's interpretation.

\section{Related Work}
\label{sec:relatedwork}

\Oca is a neuro-symbolic invariant detector, 
so we discuss prior symbolic and neural approaches.\looseness=-1

\paragraph{Augmenting Dynamic Invariant Detection}

We compare \Oca to Daikon~\cite{ErnstPGMPTX2007}, the
seminal dynamic invariant detector, which falsifies template invariants
on test-suite traces. Daikon is
the usual baseline for follow-up
work~\cite{CsallnerTS2008,MolinaDPRAF2019,MolinaPAF2021,PeiBSSY2023,MaLLXB2025}.
\Oca makes the invariants a Daikon-like system can detect more
expressive. Prior work toward this goal mainly designed new
templates~\cite{ErnstGKN99,KuzminaPGC2008,MolinaDA2022} or sped up
Daikon to evaluate more complex
invariants~\cite{ErnstCGN2000:Relevance,PerkinsE2004,GuoPME2006}.
Other approaches increase expressiveness differently.
DIG~\cite{NguyenKWF2014} infers polynomial invariants
by treating traces as high-dimensional numerical vectors; follow-ups check candidates with symbolic
execution~\cite{NguyenARH2017,NguyenDV2017} but are still limited to
numerical invariants.
Like \Oca, DIDUCE~\cite{HangalL02} checks invariants as the program runs,
but its invariants are more restrictive than Daikon's: bitmasks
recording which bits of one tracked expression
stayed constant, and their values.

\paragraph{Neuro-symbolic Dynamic Invariant Detection}

ClassInvGen~\cite{SunACTBDQL2025} uses LLMs to propose class-level
invariants for C++ data structures, not method pre-
and postconditions like \Oca. It co-generates invariants
and test inputs, discarding candidates that fail those
tests; \Oca instead uses the
project's existing test suite. ClassInvGen prompts the LLM
for side-effect-free checks but describes no mechanism for
handling side effects, nontermination, or nested evaluation during
checking, which \Oca handles. %

Balestra et al.~\cite{BalestraNMGDA2026} use LLM-generated tests to
falsify SpecFuzzer's dynamically inferred assertions, discarding
up to 12\% of the invalid ones and improving precision by up to 7\%
without affecting recall. This orthogonal work improves
the test inputs that dynamic inference relies on, not the
candidate invariants, and could be combined with \Oca to further
reduce false positives.

Pei et al.~\cite{PeiBSSY2023} fine-tuned a model on Daikon-generated
invariants to predict invariants from source code, without
running the program. Against Daikon's output on 200
executions per program, it reaches 86\% precision and recall
but outperforms Daikon only when Daikon observes five or fewer
executions. Because it imitates
Daikon, its invariants stay within Daikon's templates; \Oca instead prompts an off-the-shelf LLM
for arbitrary invariants, including calls to
project-specific methods. Their model is not public, so we could not
compare against it.

\paragraph{Neural Invariant Detection}

Pre-LLM neural approaches learned loop invariants directly. For example, Code2Inv~\cite{SiDRNS2018}
trains a graph neural network over a program graph by reinforcement
learning, with a verifier supplying the reward.
Others~\cite{RyanWYGJ2020,PadhiSM2017,FedyukovichKB2017} differ in details,
but all target inductive loop invariants over small numeric
benchmarks with a sound verifier as an oracle.
\Oca instead infers
method pre- and postconditions for real-world programs, validated
against real executions.\looseness=-1

In prior work, LLMs infer specifications and loop
invariants for program verifiers. Clause2Inv~\cite{CaoWXYWCM2025} infers
loop invariants with a generate-combine-check framework, and
SpecGen~\cite{MaLLXB2025} generates formal specifications, mutating
candidates that fail to verify;
many others pair an LLM with a verifier~\cite{MisuLMN2024,WuCYWCM2024,PirzadaRBC2024,ChakrabortyLFMLRSSS2023,LiYLDZ2026,WeiSSWWA2025,UgareSMV2026}.
Like the neural approaches above, these tools accept a candidate only if it
passes a static check, whereas \Oca accepts it if no test-suite
execution falsifies it.

Another study~\cite{EndresFCL2024} used LLMs to turn
method documentation into postconditions that would detect 64
of 525 Defects4J bugs, but
did not build an invariant detector like \Oca.

\section{Conclusion}

This paper presented \Oca, a dynamic invariant detector that obtains
candidate invariants from an LLM instead of a fixed template grammar.
Because each candidate is an arbitrary Java expression, \Oca can infer
invariants that relate many variables and call project-specific
methods, which Daikon cannot express by design.
Compared with Daikon, \Oca's invariants are more expressive, more often
valid, more useful, and cheaper to obtain. %
These results suggest that LLM-proposed
candidates can overcome the
expressiveness limits of template-based invariant detection.

\todo{The page limit is {\conferencePageLimit} pages.  This is page \thepage.}

\section*{Data Availability}
Our replication package~\cite{OcaArtifact} contains \Oca's implementation,
complete prompts, and the data and scripts to reproduce all experiments.
Upon acceptance, we will archive it on Zenodo with a DOI.

\bibliographystyle{ACM-Reference-Format}

\begin{thebibliography}{50}


\ifx \showCODEN    \undefined \def \showCODEN     #1{\unskip}     \fi
\ifx \showISBNx    \undefined \def \showISBNx     #1{\unskip}     \fi
\ifx \showISBNxiii \undefined \def \showISBNxiii  #1{\unskip}     \fi
\ifx \showISSN     \undefined \def \showISSN      #1{\unskip}     \fi
\ifx \showLCCN     \undefined \def \showLCCN      #1{\unskip}     \fi
\ifx \shownote     \undefined \def \shownote      #1{#1}          \fi
\ifx \showarticletitle \undefined \def \showarticletitle #1{#1}   \fi
\ifx \showURL      \undefined \def \showURL       {\relax}        \fi
\providecommand\bibfield[2]{#2}
\providecommand\bibinfo[2]{#2}
\providecommand\natexlab[1]{#1}
\providecommand\showeprint[2][]{arXiv:#2}

\bibitem[Alonso et~al\mbox{.}(2023)]%
        {AlonsoSR2023}
\bibfield{author}{\bibinfo{person}{Juan~C Alonso}, \bibinfo{person}{Sergio
  Segura}, {and} \bibinfo{person}{Antonio Ruiz-Cort{\'e}s}.}
  \bibinfo{year}{2023}\natexlab{}.
\newblock \showarticletitle{AGORA: automated generation of test oracles for
  REST apis}. In \bibinfo{booktitle}{\emph{International Symposium on Software
  Testing and Analysis (ISSTA)}}. \bibinfo{pages}{1018--1030}.
\newblock


\bibitem[{Anonymous}(2026)]%
        {OcaArtifact}
\bibfield{author}{\bibinfo{person}{{Anonymous}}.}
  \bibinfo{year}{2026}\natexlab{}.
\newblock \bibinfo{title}{Replication Package for ``{Template-Free Dynamic
  Invariant Detection}''}.
\newblock
  \bibinfo{howpublished}{\url{https://anonymous.4open.science/r/template-free-invariant-detection-replication-F43D/}}.
\newblock


\bibitem[{Apache Dubbo}(2026)]%
        {dubbo}
\bibfield{author}{\bibinfo{person}{{Apache Dubbo}}.}
  \bibinfo{year}{2026}\natexlab{}.
\newblock \bibinfo{title}{Apache Dubbo}.
\newblock \bibinfo{howpublished}{\url{https://github.com/apache/dubbo}}.
\newblock
\newblock
\shownote{Accessed: 2026-09-10}.


\bibitem[{Apache Hudi}(2026)]%
        {hudi}
\bibfield{author}{\bibinfo{person}{{Apache Hudi}}.}
  \bibinfo{year}{2026}\natexlab{}.
\newblock \bibinfo{title}{Apache Hudi}.
\newblock \bibinfo{howpublished}{\url{https://github.com/apache/hudi}}.
\newblock
\newblock
\shownote{Accessed: 2026-09-10}.


\bibitem[{Apollo Config}(2026)]%
        {apollo}
\bibfield{author}{\bibinfo{person}{{Apollo Config}}.}
  \bibinfo{year}{2026}\natexlab{}.
\newblock \bibinfo{title}{Apollo}.
\newblock \bibinfo{howpublished}{\url{https://github.com/apolloconfig/apollo}}.
\newblock
\newblock
\shownote{Accessed: 2026-09-10}.


\bibitem[Baker(1990)]%
        {Baker90}
\bibfield{author}{\bibinfo{person}{Henry Baker}.}
  \bibinfo{year}{1990}\natexlab{}.
\newblock \showarticletitle{Unify and Conquer (Garbage, Updating, Aliasing,
  ...)}. In \bibinfo{booktitle}{\emph{LFP '90: Proceedings of the 1990 ACM
  Conference on LISP and Functional Programming}}. \bibinfo{address}{Nice,
  France}, \bibinfo{pages}{218--226}.
\newblock


\bibitem[Balestra et~al\mbox{.}(2026)]%
        {BalestraNMGDA2026}
\bibfield{author}{\bibinfo{person}{Agust{\'\i}n Balestra},
  \bibinfo{person}{Agustin Nolasco}, \bibinfo{person}{Facundo Molina},
  \bibinfo{person}{Diego Garbervetsky}, \bibinfo{person}{Renzo Degiovanni},
  {and} \bibinfo{person}{Nazareno Aguirre}.} \bibinfo{year}{2026}\natexlab{}.
\newblock \showarticletitle{Improving Dynamic Specification Inference with
  {LLM}-Generated Counterexamples}. In \bibinfo{booktitle}{\emph{Proceedings of
  the IEEE International Conference on Software Testing, Verification and
  Validation (ICST)}}.
\newblock


\bibitem[Cao et~al\mbox{.}(2025)]%
        {CaoWXYWCM2025}
\bibfield{author}{\bibinfo{person}{Weining Cao}, \bibinfo{person}{Guangyuan
  Wu}, \bibinfo{person}{Tangzhi Xu}, \bibinfo{person}{Yuan Yao},
  \bibinfo{person}{Hengfeng Wei}, \bibinfo{person}{Taolue Chen}, {and}
  \bibinfo{person}{Xiaoxing Ma}.} \bibinfo{year}{2025}\natexlab{}.
\newblock \showarticletitle{Clause2Inv: A Generate-Combine-Check Framework for
  Loop Invariant Inference}.
\newblock \bibinfo{journal}{\emph{Proceedings of the ACM on Software
  Engineering}} \bibinfo{volume}{2}, \bibinfo{number}{ISSTA}
  (\bibinfo{year}{2025}), \bibinfo{pages}{1009--1030}.
\newblock


\bibitem[Chakraborty et~al\mbox{.}(2023)]%
        {ChakrabortyLFMLRSSS2023}
\bibfield{author}{\bibinfo{person}{Saikat Chakraborty},
  \bibinfo{person}{Shuvendu~K. Lahiri}, \bibinfo{person}{Sarah Fakhoury},
  \bibinfo{person}{Madanlal Musuvathi}, \bibinfo{person}{Akash Lal},
  \bibinfo{person}{Aseem Rastogi}, \bibinfo{person}{Aditya Senthilnathan},
  \bibinfo{person}{Rahul Sharma}, {and} \bibinfo{person}{Nikhil Swamy}.}
  \bibinfo{year}{2023}\natexlab{}.
\newblock \showarticletitle{Ranking {LLM}-Generated Loop Invariants for Program
  Verification}. In \bibinfo{booktitle}{\emph{Findings of the Association for
  Computational Linguistics: EMNLP 2023}}.
\newblock


\bibitem[Csallner et~al\mbox{.}(2008)]%
        {CsallnerTS2008}
\bibfield{author}{\bibinfo{person}{Christoph Csallner},
  \bibinfo{person}{Nikolai Tillmann}, {and} \bibinfo{person}{Yannis
  Smaragdakis}.} \bibinfo{year}{2008}\natexlab{}.
\newblock \showarticletitle{{DySy}: Dynamic symbolic execution for invariant
  inference}. In \bibinfo{booktitle}{\emph{ICSE 2008, Proceedings of the 30th
  International Conference on Software Engineering}}.
  \bibinfo{address}{Leipzig, Germany}, \bibinfo{pages}{281--290}.
\newblock
\href{https://doi.org/10.1145/1368088.1368127}{doi:\nolinkurl{10.1145/1368088.1368127}}


\bibitem[Dabic et~al\mbox{.}(2021)]%
        {seart}
\bibfield{author}{\bibinfo{person}{Ozren Dabic}, \bibinfo{person}{Emad
  Aghajani}, {and} \bibinfo{person}{Gabriele Bavota}.}
  \bibinfo{year}{2021}\natexlab{}.
\newblock \showarticletitle{Sampling projects in GitHub for MSR studies}. In
  \bibinfo{booktitle}{\emph{MSR}}. \bibinfo{pages}{560--564}.
\newblock


\bibitem[Endres et~al\mbox{.}(2024)]%
        {EndresFCL2024}
\bibfield{author}{\bibinfo{person}{Madeline Endres}, \bibinfo{person}{Sarah
  Fakhoury}, \bibinfo{person}{Saikat Chakraborty}, {and}
  \bibinfo{person}{Shuvendu~K. Lahiri}.} \bibinfo{year}{2024}\natexlab{}.
\newblock \showarticletitle{Can large language models transform natural
  language intent into formal method postconditions?}
\newblock \bibinfo{journal}{\emph{Proceedings of the ACM on Software
  Engineering}} \bibinfo{volume}{1}, \bibinfo{number}{FSE}
  (\bibinfo{year}{2024}), \bibinfo{pages}{1889--1912}.
\newblock


\bibitem[Ernst et~al\mbox{.}(2000)]%
        {ErnstCGN2000:Relevance}
\bibfield{author}{\bibinfo{person}{Michael~D. Ernst}, \bibinfo{person}{Adam
  Czeisler}, \bibinfo{person}{William~G. Griswold}, {and}
  \bibinfo{person}{David Notkin}.} \bibinfo{year}{2000}\natexlab{}.
\newblock \showarticletitle{Quickly detecting relevant program invariants}. In
  \bibinfo{booktitle}{\emph{ICSE 2000, Proceedings of the 22nd International
  Conference on Software Engineering}}. \bibinfo{address}{Limerick, Ireland},
  \bibinfo{pages}{449--458}.
\newblock
\href{https://doi.org/10.1109/icse.2000.870435}{doi:\nolinkurl{10.1109/icse.2000.870435}}


\bibitem[Ernst et~al\mbox{.}(1999)]%
        {ErnstGKN99}
\bibfield{author}{\bibinfo{person}{Michael~D. Ernst},
  \bibinfo{person}{William~G. Griswold}, \bibinfo{person}{Yoshio Kataoka},
  {and} \bibinfo{person}{David Notkin}.} \bibinfo{year}{1999}\natexlab{}.
\newblock \bibinfo{booktitle}{\emph{Dynamically discovering pointer-based
  program invariants}}.
\newblock \bibinfo{type}{{T}echnical {R}eport} UW-CSE-99-11-02.
  \bibinfo{institution}{U. Wash. Dept. of Comp. Sci. \& Eng.},
  \bibinfo{address}{Seattle, WA}.
\newblock
\newblock
\shownote{Revised Mar.~17, 2000}.


\bibitem[Ernst et~al\mbox{.}(2007)]%
        {ErnstPGMPTX2007}
\bibfield{author}{\bibinfo{person}{Michael~D. Ernst}, \bibinfo{person}{Jeff~H.
  Perkins}, \bibinfo{person}{Philip~J. Guo}, \bibinfo{person}{Stephen
  McCamant}, \bibinfo{person}{Carlos Pacheco}, \bibinfo{person}{Matthew~S.
  Tschantz}, {and} \bibinfo{person}{Chen Xiao}.}
  \bibinfo{year}{2007}\natexlab{}.
\newblock \showarticletitle{The {Daikon} system for dynamic detection of likely
  invariants}.
\newblock \bibinfo{journal}{\emph{Science of Computer Programming}}
  \bibinfo{volume}{69}, \bibinfo{number}{1--3} (\bibinfo{date}{Dec.}
  \bibinfo{year}{2007}), \bibinfo{pages}{35--45}.
\newblock
\href{https://doi.org/10.1016/j.scico.2007.01.015}{doi:\nolinkurl{10.1016/j.scico.2007.01.015}}


\bibitem[Fedyukovich et~al\mbox{.}(2017)]%
        {FedyukovichKB2017}
\bibfield{author}{\bibinfo{person}{Grigory Fedyukovich},
  \bibinfo{person}{Samuel~J. Kaufman}, {and} \bibinfo{person}{Rastislav
  Bod{\'i}k}.} \bibinfo{year}{2017}\natexlab{}.
\newblock \showarticletitle{Sampling invariants from frequency distributions}.
  In \bibinfo{booktitle}{\emph{Formal Methods in Computer-Aided Design
  (FMCAD)}}. IEEE, \bibinfo{pages}{100--107}.
\newblock


\bibitem[Guo et~al\mbox{.}(2006)]%
        {GuoPME2006}
\bibfield{author}{\bibinfo{person}{Philip~J. Guo}, \bibinfo{person}{Jeff~H.
  Perkins}, \bibinfo{person}{Stephen McCamant}, {and}
  \bibinfo{person}{Michael~D. Ernst}.} \bibinfo{year}{2006}\natexlab{}.
\newblock \showarticletitle{Dynamic inference of abstract types}. In
  \bibinfo{booktitle}{\emph{ISSTA 2006, Proceedings of the 2006 International
  Symposium on Software Testing and Analysis}}. \bibinfo{address}{Portland, ME,
  USA}, \bibinfo{pages}{255--265}.
\newblock
\href{https://doi.org/10.1145/1146238.1146268}{doi:\nolinkurl{10.1145/1146238.1146268}}


\bibitem[Hangal and Lam(2002)]%
        {HangalL02}
\bibfield{author}{\bibinfo{person}{Sudheendra Hangal} {and}
  \bibinfo{person}{Monica~S. Lam}.} \bibinfo{year}{2002}\natexlab{}.
\newblock \showarticletitle{Tracking down software bugs using automatic anomaly
  detection}. In \bibinfo{booktitle}{\emph{ICSE 2002, Proceedings of the 24th
  International Conference on Software Engineering}}.
  \bibinfo{address}{Orlando, Florida}, \bibinfo{pages}{291--301}.
\newblock
\href{https://doi.org/10.1109/icse.2002.1007976}{doi:\nolinkurl{10.1109/icse.2002.1007976}}


\bibitem[Helm et~al\mbox{.}(2018)]%
        {helm2018unified}
\bibfield{author}{\bibinfo{person}{Dominik Helm}, \bibinfo{person}{Florian
  K{\"u}bler}, \bibinfo{person}{Michael Eichberg}, \bibinfo{person}{Michael
  Reif}, {and} \bibinfo{person}{Mira Mezini}.} \bibinfo{year}{2018}\natexlab{}.
\newblock \showarticletitle{A unified lattice model and framework for purity
  analyses}. In \bibinfo{booktitle}{\emph{Proceedings of the 33rd ACM/IEEE
  International Conference on Automated Software Engineering (ASE)}}.
  \bibinfo{pages}{340--350}.
\newblock
\href{https://doi.org/10.1145/3238147.3238226}{doi:\nolinkurl{10.1145/3238147.3238226}}


\bibitem[Huang et~al\mbox{.}(2025)]%
        {huang2025prompt}
\bibfield{author}{\bibinfo{person}{Kaicheng Huang}, \bibinfo{person}{Fanyu
  Wang}, \bibinfo{person}{Yutan Huang}, {and} \bibinfo{person}{Chetan Arora}.}
  \bibinfo{year}{2025}\natexlab{}.
\newblock \showarticletitle{Prompt engineering for requirements engineering: A
  literature review and roadmap}. In \bibinfo{booktitle}{\emph{2025 IEEE 33rd
  International Requirements Engineering Conference Workshops (REW)}}. IEEE,
  \bibinfo{pages}{548--557}.
\newblock


\bibitem[Just et~al\mbox{.}(2014)]%
        {JustJE2014}
\bibfield{author}{\bibinfo{person}{Ren{\'e} Just}, \bibinfo{person}{Darioush
  Jalali}, {and} \bibinfo{person}{Michael~D. Ernst}.}
  \bibinfo{year}{2014}\natexlab{}.
\newblock \showarticletitle{{Defects4J}: A database of existing faults to
  enable controlled testing studies for {Java} programs}. In
  \bibinfo{booktitle}{\emph{ISSTA 2014, Proceedings of the 2014 International
  Symposium on Software Testing and Analysis}}. \bibinfo{address}{San Jose, CA,
  USA}, \bibinfo{pages}{437--440}.
\newblock
\href{https://doi.org/10.1145/2610384.2628055}{doi:\nolinkurl{10.1145/2610384.2628055}}
\newblock
\shownote{Tool demo}.


\bibitem[Kuzmina et~al\mbox{.}(2008)]%
        {KuzminaPGC2008}
\bibfield{author}{\bibinfo{person}{Nadya Kuzmina}, \bibinfo{person}{John Paul},
  \bibinfo{person}{Ruben Gamboa}, {and} \bibinfo{person}{James Caldwell}.}
  \bibinfo{year}{2008}\natexlab{}.
\newblock \showarticletitle{Extending dynamic constraint detection with
  disjunctive constraints}. In \bibinfo{booktitle}{\emph{WODA 2008: Workshop on
  Dynamic Analysis}}. \bibinfo{address}{Seattle, WA, USA},
  \bibinfo{pages}{57--63}.
\newblock
\href{https://doi.org/10.1145/1401827.1401839}{doi:\nolinkurl{10.1145/1401827.1401839}}


\bibitem[Lam et~al\mbox{.}(2019)]%
        {lam2019idflakies}
\bibfield{author}{\bibinfo{person}{Wing Lam}, \bibinfo{person}{Reed Oei},
  \bibinfo{person}{August Shi}, \bibinfo{person}{Darko Marinov}, {and}
  \bibinfo{person}{Tao Xie}.} \bibinfo{year}{2019}\natexlab{}.
\newblock \showarticletitle{iDFlakies: A framework for detecting and partially
  classifying flaky tests}. In \bibinfo{booktitle}{\emph{2019 12th ieee
  conference on software testing, validation and verification (icst)}}. IEEE,
  \bibinfo{pages}{312--322}.
\newblock


\bibitem[Li et~al\mbox{.}(2026)]%
        {LiYLDZ2026}
\bibfield{author}{\bibinfo{person}{Tianchi Li}, \bibinfo{person}{Zhenyu Yan},
  \bibinfo{person}{Junhao Liu}, \bibinfo{person}{Peng Di}, {and}
  \bibinfo{person}{Xin Zhang}.} \bibinfo{year}{2026}\natexlab{}.
\newblock \showarticletitle{Guiding {LLM}-based Loop Invariant Synthesis via
  Feedback on Local Reasoning Errors}.
\newblock \bibinfo{journal}{\emph{ACM Transactions on Programming Languages and
  Systems}} (\bibinfo{year}{2026}).
\newblock
\href{https://doi.org/10.1145/3806652}{doi:\nolinkurl{10.1145/3806652}}


\bibitem[{libGDX}(2026)]%
        {libgdx}
\bibfield{author}{\bibinfo{person}{{libGDX}}.} \bibinfo{year}{2026}\natexlab{}.
\newblock \bibinfo{title}{libGDX}.
\newblock \bibinfo{howpublished}{\url{https://github.com/libgdx/libgdx}}.
\newblock
\newblock
\shownote{Accessed: 2026-09-10}.


\bibitem[Ma et~al\mbox{.}(2025)]%
        {MaLLXB2025}
\bibfield{author}{\bibinfo{person}{Lezhi Ma}, \bibinfo{person}{Shangqing Liu},
  \bibinfo{person}{Yi Li}, \bibinfo{person}{Xiaofei Xie}, {and}
  \bibinfo{person}{Lei Bu}.} \bibinfo{year}{2025}\natexlab{}.
\newblock \showarticletitle{SpecGen: Automated Generation of Formal Program
  Specifications via Large Language Models}. In \bibinfo{booktitle}{\emph{2025
  IEEE/ACM 47th International Conference on Software Engineering (ICSE)}}. IEEE
  Computer Society, \bibinfo{pages}{666--666}.
\newblock


\bibitem[Misu et~al\mbox{.}(2024)]%
        {MisuLMN2024}
\bibfield{author}{\bibinfo{person}{Md~Rakib~Hossain Misu},
  \bibinfo{person}{Cristina~V. Lopes}, \bibinfo{person}{Iris Ma}, {and}
  \bibinfo{person}{James Noble}.} \bibinfo{year}{2024}\natexlab{}.
\newblock \showarticletitle{Towards {AI}-assisted synthesis of verified {Dafny}
  methods}.
\newblock \bibinfo{journal}{\emph{Proceedings of the ACM on Software
  Engineering}} \bibinfo{volume}{1}, \bibinfo{number}{FSE}
  (\bibinfo{year}{2024}).
\newblock


\bibitem[Molina et~al\mbox{.}(2022)]%
        {MolinaDA2022}
\bibfield{author}{\bibinfo{person}{Facundo Molina}, \bibinfo{person}{Marcelo
  d'Amorim}, {and} \bibinfo{person}{Nazareno Aguirre}.}
  \bibinfo{year}{2022}\natexlab{}.
\newblock \showarticletitle{Fuzzing Class Specifications}. In
  \bibinfo{booktitle}{\emph{International Conference on Software Engineering
  (ICSE)}}.
\newblock


\bibitem[Molina et~al\mbox{.}(2019)]%
        {MolinaDPRAF2019}
\bibfield{author}{\bibinfo{person}{Facundo Molina}, \bibinfo{person}{Renzo
  Degiovanni}, \bibinfo{person}{Pablo Ponzio}, \bibinfo{person}{Germ{\'a}n
  Regis}, \bibinfo{person}{Nazareno Aguirre}, {and} \bibinfo{person}{Marcelo
  Frias}.} \bibinfo{year}{2019}\natexlab{}.
\newblock \showarticletitle{Training binary classifiers as data structure
  invariants}. In \bibinfo{booktitle}{\emph{International Conference on
  Software Engineering (ICSE)}}. IEEE, \bibinfo{pages}{759--770}.
\newblock


\bibitem[Molina et~al\mbox{.}(2021)]%
        {MolinaPAF2021}
\bibfield{author}{\bibinfo{person}{Facundo Molina}, \bibinfo{person}{Pablo
  Ponzio}, \bibinfo{person}{Nazareno Aguirre}, {and} \bibinfo{person}{Marcelo
  Frias}.} \bibinfo{year}{2021}\natexlab{}.
\newblock \showarticletitle{{EvoSpex}: An Evolutionary Algorithm for Learning
  Postconditions}. In \bibinfo{booktitle}{\emph{International Conference on
  Software Engineering (ICSE)}}.
\newblock
\href{https://doi.org/10.1109/ICSE43902.2021.00112}{doi:\nolinkurl{10.1109/ICSE43902.2021.00112}}


\bibitem[Nguyen et~al\mbox{.}(2017a)]%
        {NguyenARH2017}
\bibfield{author}{\bibinfo{person}{ThanhVu Nguyen}, \bibinfo{person}{Timos
  Antonopoulos}, \bibinfo{person}{Andrew Ruef}, {and} \bibinfo{person}{Michael
  Hicks}.} \bibinfo{year}{2017}\natexlab{a}.
\newblock \showarticletitle{Counterexample-Guided Approach to Finding Numerical
  Invariants}. In \bibinfo{booktitle}{\emph{Joint Meeting on Foundations of
  Software Engineering (ESEC/FSE)}}.
\newblock


\bibitem[Nguyen et~al\mbox{.}(2017b)]%
        {NguyenDV2017}
\bibfield{author}{\bibinfo{person}{ThanhVu Nguyen}, \bibinfo{person}{Matthew~B.
  Dwyer}, {and} \bibinfo{person}{Willem Visser}.}
  \bibinfo{year}{2017}\natexlab{b}.
\newblock \showarticletitle{{SymInfer}: Inferring Program Invariants Using
  Symbolic States}. In \bibinfo{booktitle}{\emph{International Conference on
  Automated Software Engineering (ASE)}}.
\newblock


\bibitem[Nguyen et~al\mbox{.}(2014)]%
        {NguyenKWF2014}
\bibfield{author}{\bibinfo{person}{ThanhVu Nguyen}, \bibinfo{person}{Deepak
  Kapur}, \bibinfo{person}{Westley Weimer}, {and} \bibinfo{person}{Stephanie
  Forrest}.} \bibinfo{year}{2014}\natexlab{}.
\newblock \showarticletitle{DIG: A dynamic invariant generator for polynomial
  and array invariants}.
\newblock \bibinfo{journal}{\emph{ACM Transactions on Software Engineering and
  Methodology (TOSEM)}} \bibinfo{volume}{23}, \bibinfo{number}{4}
  (\bibinfo{year}{2014}), \bibinfo{pages}{1--30}.
\newblock


\bibitem[Padhi et~al\mbox{.}(2017)]%
        {PadhiSM2017}
\bibfield{author}{\bibinfo{person}{Saswat Padhi}, \bibinfo{person}{Rahul
  Sharma}, {and} \bibinfo{person}{Todd Millstein}.}
  \bibinfo{year}{2017}\natexlab{}.
\newblock \showarticletitle{{LoopInvGen}: A loop invariant generator based on
  precondition inference}.
\newblock \bibinfo{journal}{\emph{arXiv preprint arXiv:1707.02029}}
  (\bibinfo{year}{2017}).
\newblock


\bibitem[Pei et~al\mbox{.}(2023)]%
        {PeiBSSY2023}
\bibfield{author}{\bibinfo{person}{Kexin Pei}, \bibinfo{person}{David Bieber},
  \bibinfo{person}{Kensen Shi}, \bibinfo{person}{Charles Sutton}, {and}
  \bibinfo{person}{Pengcheng Yin}.} \bibinfo{year}{2023}\natexlab{}.
\newblock \showarticletitle{Can large language models reason about program
  invariants?}. In \bibinfo{booktitle}{\emph{International Conference on
  Machine Learning}}. PMLR, \bibinfo{pages}{27496--27520}.
\newblock


\bibitem[Perkins and Ernst(2004)]%
        {PerkinsE2004}
\bibfield{author}{\bibinfo{person}{Jeff~H. Perkins} {and}
  \bibinfo{person}{Michael~D. Ernst}.} \bibinfo{year}{2004}\natexlab{}.
\newblock \showarticletitle{Efficient incremental algorithms for dynamic
  detection of likely invariants}. In \bibinfo{booktitle}{\emph{FSE 2004:
  Proceedings of the {ACM} {SIGSOFT} 12th Symposium on the Foundations of
  Software Engineering}}. \bibinfo{address}{Newport Beach, CA, USA},
  \bibinfo{pages}{23--32}.
\newblock
\href{https://doi.org/10.1145/1029894.1029901}{doi:\nolinkurl{10.1145/1029894.1029901}}


\bibitem[Pirzada et~al\mbox{.}(2024)]%
        {PirzadaRBC2024}
\bibfield{author}{\bibinfo{person}{Muhammad A.~A. Pirzada},
  \bibinfo{person}{Giles Reger}, \bibinfo{person}{Ahmed Bhayat}, {and}
  \bibinfo{person}{Lucas~C. Cordeiro}.} \bibinfo{year}{2024}\natexlab{}.
\newblock \showarticletitle{{LLM}-Generated Invariants for Bounded Model
  Checking Without Loop Unrolling}. In \bibinfo{booktitle}{\emph{Proceedings of
  the 39th IEEE/ACM International Conference on Automated Software Engineering
  (ASE)}}.
\newblock


\bibitem[Ryan et~al\mbox{.}(2020)]%
        {RyanWYGJ2020}
\bibfield{author}{\bibinfo{person}{Gabriel Ryan}, \bibinfo{person}{Justin
  Wong}, \bibinfo{person}{Jianan Yao}, \bibinfo{person}{Ronghui Gu}, {and}
  \bibinfo{person}{Suman Jana}.} \bibinfo{year}{2020}\natexlab{}.
\newblock \showarticletitle{{CLN2INV}: Learning loop invariants with continuous
  logic networks}. In \bibinfo{booktitle}{\emph{International Conference on
  Learning Representations (ICLR)}}.
\newblock


\bibitem[Sahoo et~al\mbox{.}(2024)]%
        {sahoo2024systematic}
\bibfield{author}{\bibinfo{person}{Pranab Sahoo}, \bibinfo{person}{Ayush~Kumar
  Singh}, \bibinfo{person}{Sriparna Saha}, \bibinfo{person}{Vinija Jain},
  \bibinfo{person}{Samrat Mondal}, {and} \bibinfo{person}{Aman Chadha}.}
  \bibinfo{year}{2024}\natexlab{}.
\newblock \showarticletitle{A systematic survey of prompt engineering in large
  language models: Techniques and applications}.
\newblock \bibinfo{journal}{\emph{arXiv preprint arXiv:2402.07927}}
  (\bibinfo{year}{2024}).
\newblock


\bibitem[Schulhoff et~al\mbox{.}(2025)]%
        {schulhoff2025prompt}
\bibfield{author}{\bibinfo{person}{S Schulhoff}, \bibinfo{person}{M Ilie},
  \bibinfo{person}{N Balepur}, \bibinfo{person}{K Kahadze}, \bibinfo{person}{A
  Liu}, \bibinfo{person}{C Si}, \bibinfo{person}{Y Li}, \bibinfo{person}{A
  Gupta}, \bibinfo{person}{H Han}, \bibinfo{person}{S Schulhoff},
  {et~al\mbox{.}}} \bibinfo{year}{2025}\natexlab{}.
\newblock \bibinfo{title}{The prompt report: A systematic survey of prompt
  engineering techniques (arXiv: 2406.06608). arXiv}.
\newblock


\bibitem[Si et~al\mbox{.}(2018)]%
        {SiDRNS2018}
\bibfield{author}{\bibinfo{person}{Xujie Si}, \bibinfo{person}{Hanjun Dai},
  \bibinfo{person}{Mukund Raghothaman}, \bibinfo{person}{Mayur Naik}, {and}
  \bibinfo{person}{Le Song}.} \bibinfo{year}{2018}\natexlab{}.
\newblock \showarticletitle{Learning loop invariants for program verification}.
  In \bibinfo{booktitle}{\emph{Advances in Neural Information Processing
  Systems (NeurIPS)}}.
\newblock


\bibitem[{Spring}(2026)]%
        {spring}
\bibfield{author}{\bibinfo{person}{{Spring}}.} \bibinfo{year}{2026}\natexlab{}.
\newblock \bibinfo{title}{Spring Framework}.
\newblock
  \bibinfo{howpublished}{\url{https://github.com/spring-projects/spring-framework}}.
\newblock
\newblock
\shownote{Accessed: 2026-09-10}.


\bibitem[Sun et~al\mbox{.}(2025)]%
        {SunACTBDQL2025}
\bibfield{author}{\bibinfo{person}{Chuyue Sun}, \bibinfo{person}{Viraj Agashe},
  \bibinfo{person}{Saikat Chakraborty}, \bibinfo{person}{Jubi Taneja},
  \bibinfo{person}{Clark Barrett}, \bibinfo{person}{David Dill},
  \bibinfo{person}{Xiaokang Qiu}, {and} \bibinfo{person}{Shuvendu~K Lahiri}.}
  \bibinfo{year}{2025}\natexlab{}.
\newblock \showarticletitle{ClassInvGen: Class invariant synthesis using large
  language models}. In \bibinfo{booktitle}{\emph{International Symposium on AI
  Verification}}. Springer, \bibinfo{pages}{64--96}.
\newblock


\bibitem[{The Netty Project}(2026)]%
        {netty}
\bibfield{author}{\bibinfo{person}{{The Netty Project}}.}
  \bibinfo{year}{2026}\natexlab{}.
\newblock \bibinfo{title}{Netty}.
\newblock \bibinfo{howpublished}{\url{https://github.com/netty/netty}}.
\newblock
\newblock
\shownote{Accessed: 2026-08-28}.


\bibitem[Ugare et~al\mbox{.}(2026)]%
        {UgareSMV2026}
\bibfield{author}{\bibinfo{person}{Shubham Ugare}, \bibinfo{person}{Tarun
  Suresh}, \bibinfo{person}{Sasa Misailovic}, {and} \bibinfo{person}{Julien
  Vanegue}.} \bibinfo{year}{2026}\natexlab{}.
\newblock \showarticletitle{{Fun2spec}: Code Contract Synthesis At Scale}. In
  \bibinfo{booktitle}{\emph{Companion Proceedings of the ACM International
  Conference on the Foundations of Software Engineering (FSE), Industry
  Papers}}.
\newblock


\bibitem[Wei et~al\mbox{.}(2025)]%
        {WeiSSWWA2025}
\bibfield{author}{\bibinfo{person}{Anjiang Wei}, \bibinfo{person}{Tarun
  Suresh}, \bibinfo{person}{Tianran Sun}, \bibinfo{person}{Haoze Wu},
  \bibinfo{person}{Ke Wang}, {and} \bibinfo{person}{Alex Aiken}.}
  \bibinfo{year}{2025}\natexlab{}.
\newblock \bibinfo{title}{{InvBench}: Can {LLMs} Accelerate Program
  Verification with Invariant Synthesis?}
\newblock \bibinfo{howpublished}{arXiv:2509.21629}.
\newblock


\bibitem[Wu et~al\mbox{.}(2024)]%
        {WuCYWCM2024}
\bibfield{author}{\bibinfo{person}{Guangyuan Wu}, \bibinfo{person}{Weining
  Cao}, \bibinfo{person}{Yuan Yao}, \bibinfo{person}{Hengfeng Wei},
  \bibinfo{person}{Taolue Chen}, {and} \bibinfo{person}{Xiaoxing Ma}.}
  \bibinfo{year}{2024}\natexlab{}.
\newblock \showarticletitle{{LLM} Meets Bounded Model Checking: Neuro-symbolic
  Loop Invariant Inference}. In \bibinfo{booktitle}{\emph{Proceedings of the
  39th IEEE/ACM International Conference on Automated Software Engineering
  (ASE)}}.
\newblock


\bibitem[Xia et~al\mbox{.}(2025)]%
        {XiaSPDRR2025}
\bibfield{author}{\bibinfo{person}{Yuan Xia}, \bibinfo{person}{Deepayan Sur},
  \bibinfo{person}{Aabha~Shailesh Pingle}, \bibinfo{person}{Jyotirmoy~V
  Deshmukh}, \bibinfo{person}{Mukund Raghothaman}, {and}
  \bibinfo{person}{Srivatsan Ravi}.} \bibinfo{year}{2025}\natexlab{}.
\newblock \showarticletitle{Discovering Likely Invariants for Distributed
  Systems Through Runtime Monitoring and Learning}. In
  \bibinfo{booktitle}{\emph{International Conference on Verification, Model
  Checking, and Abstract Interpretation}}. Springer, \bibinfo{pages}{3--25}.
\newblock


\bibitem[Yuan et~al\mbox{.}(2025)]%
        {yuan2025designrepair}
\bibfield{author}{\bibinfo{person}{Mingyue Yuan}, \bibinfo{person}{Jieshan
  Chen}, \bibinfo{person}{Zhenchang Xing}, \bibinfo{person}{Aaron Quigley},
  \bibinfo{person}{Yuyu Luo}, \bibinfo{person}{Tianqi Luo},
  \bibinfo{person}{Gelareh Mohammadi}, \bibinfo{person}{Qinghua Lu}, {and}
  \bibinfo{person}{Liming Zhu}.} \bibinfo{year}{2025}\natexlab{}.
\newblock \showarticletitle{{DesignRepair}: Dual-stream design guideline-aware
  frontend repair with large language models}. In
  \bibinfo{booktitle}{\emph{2025 IEEE/ACM 47th International Conference on
  Software Engineering (ICSE)}}. IEEE, \bibinfo{pages}{2483--2494}.
\newblock


\bibitem[Zeller and Hildebrandt(2002)]%
        {zeller2002simplifying}
\bibfield{author}{\bibinfo{person}{Andreas Zeller} {and} \bibinfo{person}{Ralf
  Hildebrandt}.} \bibinfo{year}{2002}\natexlab{}.
\newblock \showarticletitle{Simplifying and Isolating Failure-Inducing Input}.
\newblock \bibinfo{journal}{\emph{IEEE Transactions on Software Engineering}}
  \bibinfo{volume}{28} (\bibinfo{year}{2002}), \bibinfo{pages}{183--200}.
\newblock
\href{https://doi.org/10.1109/32.988498}{doi:\nolinkurl{10.1109/32.988498}}


\end{thebibliography}

%
%

\end{document}